\documentclass[ aps, rsi, amsmath, amssymb, preprint]{revtex4-2}

\usepackage{graphicx}
\usepackage{bm}

\usepackage{hyperref}

\usepackage{color}

\begin{document}


\newcommand{\be}{\begin{equation}} 
\newcommand{\ee}{\end{equation}}
\newcommand{\bea}{\begin{eqnarray}}   
\newcommand{\eea}{\end{eqnarray}}

\newcommand{\rr}{{\bf r}}

\newcommand{\nvec}{\boldsymbol{n}}
\newcommand{\txi}{\boldsymbol{\xi}^t}
\newcommand{\F}{\boldsymbol{F}}
\newcommand{\colxi}{\boldsymbol{u}}

\newcommand{\UU}{{\cal U}}
\newcommand{\xchi}{\boldsymbol{\chi}}
\newcommand{\pvec}{\boldsymbol{p}}
\newcommand{\eeta}{\boldsymbol{\eta}}
\newcommand{\xxi}{\boldsymbol{\xi}}

\newcommand{\xx}{\boldsymbol{x}}
\newcommand{\vv}{\boldsymbol{v}}
\newcommand{\uu}{\boldsymbol{u}}
\newcommand{\xib}{\boldsymbol{\xi}}

\date{\today}
\title{Active Gaussian Network Model: a non-equilibrium description of protein fluctuations and allosteric behavior }

\author{Giulio Costantini}
\affiliation{Institute for Complex Systems - CNR - P.le Aldo Moro 2 - Rome - Italy.}

\author{Lorenzo Caprini}
\affiliation{Sapienza University of Rome - P.le Aldo Moro 5 - Rome - Italy.}

\author{Umberto Marini Bettolo Marconi}
\affiliation{University of Camerino - Via Madonna delle Carceri - Camerino - Italy.}

\author{Fabio Cecconi}
\affiliation{CNR-Istituto dei Sistemi Complessi Via dei Taurini 19 - Rome - Italy and INFN Sez. di Roma; Sapienza University of Rome - Piazzale A. Moro - Rome - Italy.}

\begin{abstract}
Understanding the link between structure and function in proteins is fundamental in molecular biology and proteomics.
A central question in this context is whether allostery - where the binding of a molecule at one site affects the activity of a distant site - emerges as a further manifestation of the intricate interplay between structure, function, and intrinsic dynamics.
This study explores how allosteric regulation is modified when intrinsic protein dynamics operate under out-of-equilibrium conditions.
To this purpose, we introduce a simple nonequilibrium model of protein dynamics, inspired by active matter systems, by generalizing the widely employed Gaussian Network Model (GNM) to incorporate non-thermal effects. 
Our approach underscores the advantage of framing allostery as a causal process by using, as a benchmark system, the second PDZ domain of the human phosphatase hPT1E that mediates protein-protein interactions.
We employ causal indicators, such as response functions and transfer entropy, to identify the network of PDZ2 residues through which the allosteric 
signal propagates across the protein structure. These indicators reveal specific regions that align well with experimental observations. Furthermore, our results suggest that deviations from purely thermal fluctuations can significantly influence allosteric communication by introducing distinct timescales and memory effects. 
This influence is particularly relevant when the allosteric response unfolds on timescales incompatible with relaxation to equilibrium.
Accordingly, non-thermal fluctuations may become essential for accurately describing protein responses to ligand binding and developing a comprehensive understanding of allosteric regulation.
\end{abstract}

\date{\today}	
\pacs{Valid PACS appear here}
\keywords{Suggested keywords}

\maketitle

\section{Introduction}
Many proteins to perform their functions undergo conformational changes or dynamical shifts involving structural fluctuations and alterations in flexibility~\cite{lodish2002molecular}. 
These transitions, often driven by non-equilibrium processes under physiological conditions, are crucial for regulating protein activity and signalling~\cite{smock2009sending}. 
Increasing evidence and a certain consensus~\cite{kubitzkiChapter12} agree that protein function is not merely assisted but may be directly mediated by structural fluctuations. Rather than being random, these fluctuations are dictated and organized by the geometry and topology of the native fold, highlighting the deep interdependence of  
\begin{equation}
\mathrm{Structure} - \mathrm{Dynamics} - \mathrm{Function}.
\label{eq:StrDynFunk}
\end{equation}  
This perspective extends the classical structure-function paradigm~\cite{lesk2010introduction,mclaughlin2012spatial} into a more general framework that incorporates protein dynamics~\cite{smock2009sending,kesselBook_2018}.
Therefore, proteins are not acknowledged as rigid scaffolds but as complex ``machines'' that, through their dynamics, process information from the environment to sample an ensemble of conformations that facilitate their function~\cite{bahar2010global,erman2025fluctuation}.
In theoretical approaches, the relationship~\eqref{eq:StrDynFunk} is frequently invoked to infer function through Normal Mode Analysis (NMA) or Principal Component Analysis (PCA) through the notion of functional modes~\cite{hub2009detection,dykeman2010normal,kubitzkiChapter12}.  
Such modes correspond to low-frequency and large-scale conformational changes able to facilitate or mediate key biological activity by modulating the intrinsic dynamics of the protein.

The open question remains whether the structure-dynamics-function relationship also underlies one of the most intriguing biological processes, referred to as allosterism ~\cite{frauenfelder2001role,liu2016allostery,thirumalai2019symmetry}. 
Allostery is a chemical regulation process~\cite{ribeiro2016chemical} that occurs in several enzymes and proteins where binding of a ligand to a site (referred to as the regulation site) enables activation of a distal site (active site), so that the macromolecule is ``ready'' to perform its biological function.
It is a sort of long-range communication established between the ligand-binding site and the active region~\cite{monod1961general,guarnera2016remote}. 
In the traditional view, allosteric processes are mainly attributed to conformational changes, however
Cooper and Dryden~\cite{cooper1984allostery}, proposed the allostery as a result of conformational dynamics, summarized by the well-known proposition ``allostery without conformational changes''. 
In this alternative view, allosteric communication is governed or modulated by intrinsic dynamical fluctuations, positioning allosteric regulation in proteins as a further manifestation of paradigm~\ref{eq:StrDynFunk}.

A popular approach to exploring allostery involves NMA and PCA within covariance analysis to identify the coordinated motions of different residues, potentially revealing allosteric association among residues~\cite{allostery_methods} mediated and modulated by functional modes~\cite{guarnera2016str}.
The alternative to expensive full-atomistic NMA is the coarse-grained surrogate Elastic Network Model~\cite{tirion1996,sanejouand2013elastic,ENM}
that replaces the complexity of the aminoacids with nodes connected by springs (replacing real interactions) according to a simple proximity or geometrical criterion.
This drastic reduction of the protein to a simplified system remains realistic as long as it correctly embodies the topology and structural properties of the native fold \cite{micheletti2004accurate,micheletti2002crucial}.
The isotropic version of the ENM is known as the Gaussian Network Model (GNM), because it treats fluctuations as scalar deviations rather than full 3D displacement vector~\cite{GNM_1997}.
The main advantage of these models lies in their straightforward numerical implementation and exact solvability via matrix diagonalization.

In this study, we introduce a modified version of the traditional GNM, referred to as the active GNM (aGNM), by analogy with active systems. In the aGNM, the thermal bath is replaced by a colored-noise bath that captures out-of-equilibrium fluctuations with memory effects while retaining analytical tractability.
Through this minimal and solvable model, we try to address the role of out-of-equilibrium dynamics in allosteric mechanisms~\cite{austin1975dynamics,stockPDZ3_nonEQ,stock2018non,timeResPDZ,xingNEQallo}. 
As a test system, we focus on the PDZ2 domain, which is known for its well-characterized allosteric properties.
In this context, our approach seeks to understand how deviations from thermal fluctuations influence allosteric regulation, taking into account that several allosteric responses are incompatible with pure thermodynamic models~\cite{hathcock_NEQallo}.

Deviations from pure thermal behavior can be expected at a molecular level, because a protein, during its allosteric response, 
does not necessarily undergo a fast-enough reorganization to explore its entire ensemble of accessible conformations.  
In practice, allosteric communication often occurs on timescales where conformational dynamics cannot be expected to be fully equilibrated~\cite{Goodey2008,Motlagh2014,cecconiPDZ_diff}. 
This condition challenges traditional equilibrium models and supports the growing evidence that non-equilibrium dynamics play a crucial role in allosteric regulation.

Moreover, proteins and enzymes frequently operate in energy-driven environments or under chemical gradients, conditions that cannot be reduced to a thermal bath alone but necessitate other sources of fluctuation characterized by an intrinsic memory~\cite{barducci2015non}.
As an example, proteins interact with a structured hydration layer, in which water molecules display collective dynamics and a relaxation slower than bulk water. These interactions may introduce long-lived correlations that lead to deviations from the assumptions of a simple thermal bath.
In this respect, the aGNM, by incorporating non-equilibrium effects, provides a minimal framework for generalizing protein dynamical fluctuations beyond thermal equilibrium.

In our approach to PDZ2 allostery, we adopt a causal perspective, where the binding of the effector at the controlling site is considered the "cause", and the resulting allosteric response at the active site is the "effect." 
This interpretation allows us to employ causal indicators as tools to identify and quantify relationships among PDZ2 residues, 
providing a systematic way to study allosteric communication.

The paper is organized as follows. 
In Section~\ref{sec:pdz}, we briefly review PDZ allostery as reported in the literature to establish the biological context of the problem.
Section~\ref{sec:model} describes the classical coarse-grained approach to protein thermal fluctuations to which we added active-like effects, to define aGNM, emulating non-equilibrium condition without spoiling the solvability of the model.
In Section~\ref{sec:valida}, we validate the aGNM by comparing the correlation patterns predicted by our model with experimental data.
Section~\ref{sec:causal} presents the aGNM results for PDZ2 allostery, analyzed through a causality-based approach using causal indicators specifically developed for the aGNM framework.
Finally, section~\ref{sec:concl} is devoted to discussion and conclusions.

\section{Short summary of PDZ2 allostery
\label{sec:pdz}}
Before proceeding with modeling and results, it is convenient to discuss some key features 
of {\em allosterism in the PDZ family}~\cite{PDZallofamily}.
Here, we consider the second PDZ domain (PDZ2) from human Protein Tyrosine Phosphatase 1E (hPTP1E), 
a well-studied member of the PDZ family~\cite{PDZallofamily}.
The name PDZ refers to a domain found in certain proteins, whose name comes 
from the first three proteins where this domain was first identified:
Postsynaptic density protein (PSD95), Drosophila tumor suppressor protein (DlgA) and Zonula occludens-1 protein (ZO-1).

PDZ domains are important for protein-protein interactions and are involved in organizing signaling complexes at cell membranes.
While the PDZ family exhibits a relatively low sequence identity, it maintains a strong affinity for its characteristic secondary structure, strongly suggesting that native state topology may play a role in determining the allosteric behavior.
Lockless and Ranganathan~\cite{lockless_ranganath99}, by using statistical coupling analysis, revealed coevolving residue networks in the PDZ domain, highlighting structural and functional connectivity in proteins.

\begin{figure*}[t!]
\centering
\includegraphics[width=\textwidth]{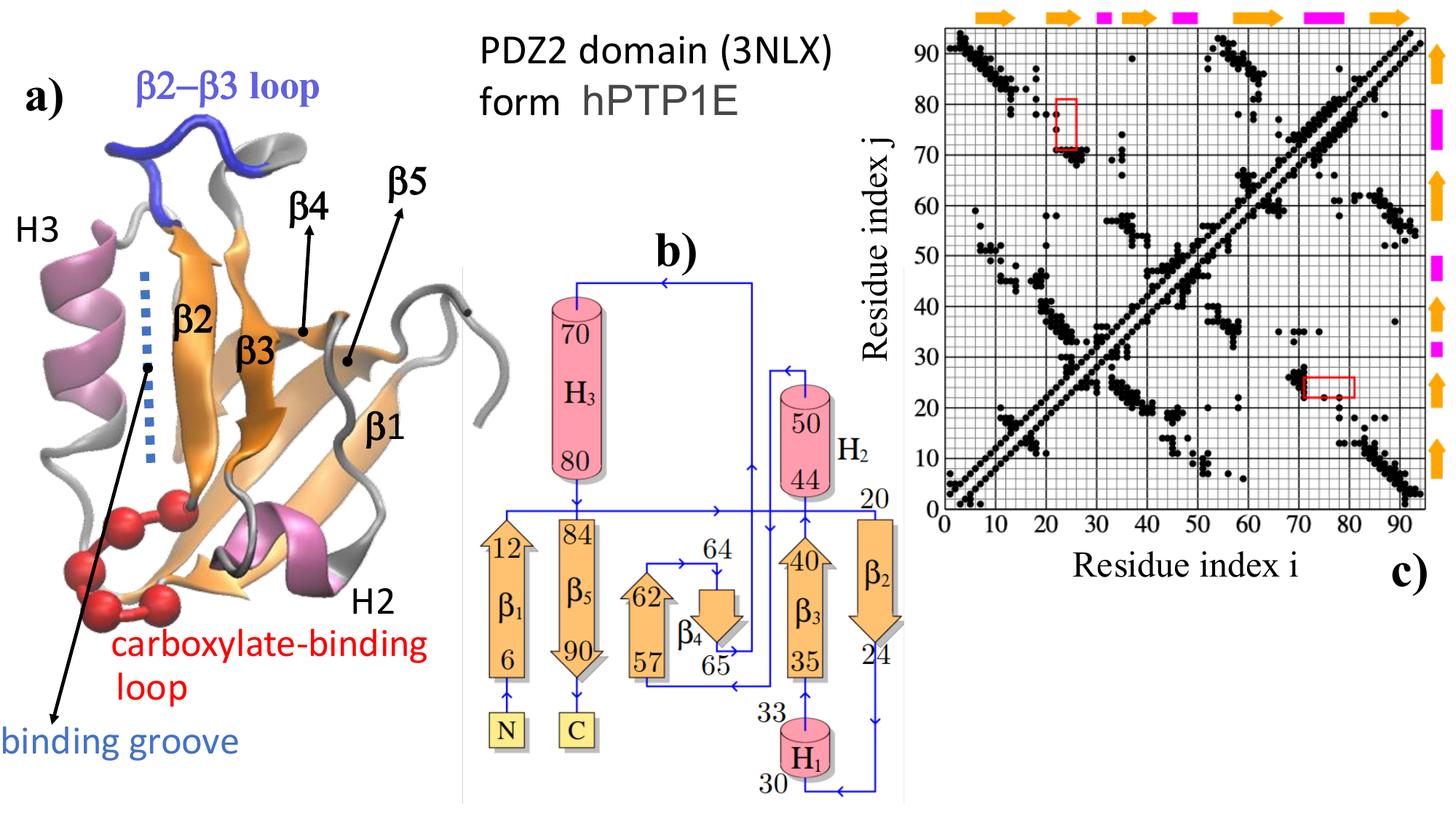}
\caption{
Structure and topology of the PDZ2 analyzed in this work (PDBid: 3NLX). 
Panel (a): Cartoon representation of the PDZ2 highlighting in red balls
the carboxylate-binding loop ($\beta$1-$\beta$2) and in blue the loop between ($\beta$2-$\beta$3) comprising the small 3-10 helix H2.
Panel (b): Topology plot of PDZ2 showing the secondary structure content with five beta sheets and three helices. 
Panel (c): Contact map defined by the heavy-atom contacts using a cutoff of $R_c=5$\AA; 
the cutoff identifies $371$ non-bonded native contacts, i.e. such that $|i-j|\ge 2$. The secondary-structure motives are reported at the 
borders to spot the direct correspondence between interactions and structure.
\label{fig:3LNX}
}
\end{figure*}

The PDZ structure we analyze is represented in panel (a) of Fig.\ref{fig:3LNX} and it is extracted from the PDB-file 3LNX; 
it is composed of 5 $\beta$-strands, two $\alpha$-helices and a 3-10 helix, precisely: 
$\beta$1 (residues 6-12), $\beta$2 (20-24), $H$1 (30-33),  
$\beta$3 (35-40), $H$2 (44-50), $\beta$4 (57-65), $H$3 (70-80), and $\beta$5 (84-90), see Fig.~\ref{fig:3LNX}~(b).
Notice that, in contrast to 3LNX PDB classification, we consider as a unique $\beta$ strand, the segment $57-65$ 
to simplify the analysis.

Allosteric behavior is known to be triggered by a ligand binding motif into the hydrophobic binding groove hosted between the $H$3-helix and the $\beta$2-strand~\cite{alloPDZatomic}.
The final three to five C-terminal residues of PDZ domains interact with target proteins through the carboxylate-binding loop.
This loop facilitates binding to the C-terminal carboxylate group of the target protein, and it is characterized by a conserved X-$\Phi$-Gly-$\Phi$ sequence, where X is a generic amino acid and $\Phi$ indicates a hydrophobic residue~\cite{carboxylate}.  

In summary, the PDZ2 (3LNX) allosteric network is primarily defined by a network of dynamically communicating residues rather than distinct structural sites.
Based on computational and experimental studies~\cite{kozlov2002solution,ivarsson2012plasticity}, key structural elements involved in allosteric regulation include~\cite{PDZallofamily}:
\begin{itemize}
\item binding groove limited by $\beta$2 and helix H3;
\item loop region $\beta$2-$\beta$3 including the small 3-10 helix H1. This loop closes the ligand-binding groove from above 
in Fig.\ref{fig:3LNX}(a);
\item H2 helix, important for intramolecular signaling;
\item The carboxylate-binding loop $\beta$1-$\beta$2, which stabilizes the C-terminal carboxylate of target peptides through hydrogen bonding and electrostatic interactions, ensuring high-affinity and specificity in PDZ-target interactions.
\end{itemize}
Therefore, the residues involved in PDZ2 allostery are typically located within and around its binding groove and neighboring secondary 
structural elements, which allow for conformational changes. 
In summary, key residues often include those in the $H$2 helix, $\beta$2 strand, and the binding groove, as these areas contribute to ligand binding and can transmit allosteric signals.

\section{A simple model for protein fluctuations in non-equilibrium environments
\label{sec:model}}
We introduce a new model, the active Gaussian Network Model (aGNM), to describe the protein fluctuation dynamics in non-equilibrium environments. 
The model includes in the standard framework of GNM \cite{GNM_1997,Bahar_FoldDes} the intrinsic activity that governs the dynamics of active matter systems~\cite{bechinger2016active,digregorio2018full,caprini2020spontaneous}. 
Although proteins cannot be considered self-propelled objects, their fundamental units experience non-equilibrium fluctuations, particularly in the cell environment, where continuous energy consumption, crowding, and chemical gradients produce fluctuations expected to deviate from thermal fluctuations.
Here, this deviation, which we call activity in analogy with active systems, is exemplified through a noise term with memory, 
by employing the active Ornstein-Uhlenbeck dynamics~\cite{szamel2014self,maggi2014generalized,caprini2018active}.
In our description, activity does not lead to structural changes in the protein, rather generates non-equilibrium fluctuations that affect 
properties and statistics of observables.

In our coarse-grained approach, individual aminoacids of the PDZ2 are identified with their $\alpha$-carbon atoms, C$_{\alpha}$, in the native fold, as obtained from PDBid: 3LNX. 
The protein is then portrayed as a mechanical network where nodes correspond to the C$_{\alpha}$ positions and inter-residue 
interactions are replaced by Hookean springs.
The system's potential energy is given by:
\begin{equation}
V_{GNM} = \frac{g}{2} \sum_{i,j} K_{ij}\;
\Delta\mathbf{r}_i \cdot \Delta\mathbf{r}_j\;,
\label{eq:GNM}
\end{equation}
where, as explained below, $K_{ij}$ encodes the connectivity of the network, and $\Delta\mathbf{r}_i$ are the displacements from the 
native positions of C$_{\alpha}$ taken as equilibrium positions.
In the following, we will redefine the displacements $\Delta\mathbf{r}_i \to \mathbf{r}_i$ to simplify the notation.
This GNM-based framework provides a simplified yet nontrivial description of PDZ2 as long as the network of harmonic interactions 
correctly encodes the topology of its native fold.
Specifically, the factor $g$ (dimension of Energy/Length$^2$) is a scale factor of the harmonic potential that can be set by matching the predicted 
mean-square-displacement (MSD) of C$_{\alpha}$'s from their native positions with the experimental crystallographic 
B-factors~\cite{bfactors,burioni04,GNM_vs_Exper}.
The coefficients $K_{ij}$ are the elements of the coupling matrix $\mathbb{K}$, often termed Kirchhoff's matrix defined through the weighted contact matrix elements, $A_{ij}$ (weighted adjaciency matrix of the C${_\alpha}$ network) via the relation  
\begin{equation}
\mathbb{K}_{ij} = 
\begin{cases}
- A_0                  & |i-j| = 1 \\
- A_{ij}               & |i-j| > 1\\
\sum_{l=1,N} A_{il}    &  i=j  \;, 
\end{cases}
\end{equation}
where $A_0$ is the strength of the harmonic interaction along the chain (backbone).  
To distinguish the role of the backbone links from the rest of the network, we set $A_0\sim 10 A_{ij}$ and $A_{ij}=1$ \cite{ceccoPhysBio}.  

We consider the GNM based on a ``heavy-atom contact-map'' (briefly heavy-map) \cite{hmap}, such that a couple of residues $i-j$ is 
connected by a spring (thus $A_{ij}>0$), if in the native state of PDZ2 (3LNX), 
the aminoacids $i$ and $j$ have at least a couple of heavy atoms a1, a2 in contact, i.e. if their distance $r_{\mathrm{a1},\mathrm{a2}}\le r_c = 5$\AA.
The obtained heavy-map, representing the native interactions, is displayed in Fig.~\eqref{fig:3LNX}~(c).

\subsection{The Active Gaussian Network Model}
Activity in the above GNM is introduced via an active velocity, $w_i$, evolving as an Ornstein-Uhlenbeck process  
with typical autocorrelation time $\tau_a$ and variance $v_a^2 = k_B T_a/m$. 
This model for activity is well-known in active matter~\cite{bechinger2016active,szamel2014self, maggi2014generalized, caprini2018active} and has the advantage of allowing analytical treatments for liquids consisting of active particles as well as active matter collective phenomena~\cite{digregorio2018full,mandal2019motility, caprini2020spontaneous,marconi2015towards,caprini2020time}.
The autocorrelation time $\tau_a$ of the OU process can be identified as the leading timescale characterizing the memory induced by the non-equilibrium environment.

This system, which we call the active Gaussian Network model (aGNM), is governed by a set of overdamped stochastic equations for the displacement $x_i$ and active velocity $w_i$:
\begin{align}
\gamma \dot{x}_i &= - g \sum_j \mathbb{K}_{ij} x_j + \gamma w_i + \sqrt{\dfrac{2 \gamma k_B T_0}{m}}\;\xi_i(t)
\label{eq:evol1}\\ 
       \dot{w}_i &= - \dfrac{w_i}{\tau_a} + \sqrt{\dfrac{2 k_B T_a}{ m\tau_a}}\;\eta_i(t) \,.
\label{eq:evol2}
\end{align}
Here, $T_0$ represents the heat-bath temperature, $k_B$ denotes the Boltzmann constant while $\gamma$ represents the friction coefficient in unit of mass - thus $\gamma$ is an inverse of time.
The mass $m$ is taken to be the average mass of the residues in the 3LNX structure: 
$m=10.685\,\mbox{kDa}/94 \simeq 1.89\times10^{-25}$kg.
Despite the notation, $T_a$ does not represent the system temperature but quantifies the strength of the nonequilibrium 
fluctuations, while $\tau_a$ corresponds to the typical memory time ({\em i.e.} the persistence or memory time).
Finally, $\xi_i,\eta_i$ are two independent zero-average and time delta-correlated Gaussian processes. 

The active aGNM is fully solvable as it defines a multivariate Ornstein-Uhlenbeck process, requiring only numerical diagonalization 
of the sparse matrix \(\mathbb{K}\) that can be expressed as:  
$$
\mathbb{K} = \mathbb{U} \Lambda \mathbb{U}^{\dagger}
$$
where $\Lambda$ is a diagonal matrix containing eigenvalues $\{\lambda_1=0, \lambda_2, \dots, \lambda_N\}$, and $\mathbb{U}$ 
is the matrix with orthogonal eigenvectors as columns, such that $\mathbb{U} \mathbb{U}^{\dagger} = \mathbb{I}$, being $\mathbb{I}$ the identity matrix.

To solve Eqs.~(\ref{eq:evol1},\ref{eq:evol2}), we expand positions $\mathbf{x}$, activities $\mathbf w$, and thermal noise $\boldsymbol{\xi}$ in the base of eigenmodes $\{\mathbf{u}(1),\ldots,\mathbf{u}(N)\}$ of $\mathbb{K}$, as
\begin{subequations}
    \begin{align}
        \label{eq:x(t)_solution}
        &\mathbf{x}(t) = \sum_k Q_k(t)\,\mathbf{u}(k)\\
        &\mathbf{w}(t) = \sum_k P_k(t)\,\mathbf{u}(k)\\
        &\boldsymbol{\xi}(t) = \sum_k \phi_k(t)\,\mathbf{u}(k)\;\,.
    \end{align}
\end{subequations}
By substituting these expressions in the dynamics~(\ref{eq:evol1},\ref{eq:evol2}), we obtain an equivalent equation for $Q_k$, $P_k$ and $\psi_k$ 
\begin{equation}
\gamma \dot{Q}_k(t) = -g\,\lambda_k\ Q_k(t) + \gamma\,P_k(t) + \sqrt{\dfrac{2 \gamma k_B T_0}{m}}\;\phi_k(t) \,.
\label{eq:dotQk}
\end{equation}
The stationary solution of Eq.~\eqref{eq:dotQk} reads
\begin{equation}
Q_k(t) = e^{-\mu_k t} 
\int_{-\infty}^t\! ds\;e^{\mu_k s}\;\bigg[P_k(s) + \sqrt{\dfrac{2 k_B T_0}{m\gamma}}\;\phi_k(s)\bigg] \,,
\label{eq:solQk}
\end{equation}
where $\mu_k=g\lambda_k/\gamma$.
The solution for the coordinate $x_i$ can be obtained by combining Eq.~\eqref{eq:x(t)_solution} with the solution~Eq.\eqref{eq:solQk}
\begin{equation}
x_i(t) = \sum_k Q_k(t)\;u_i(k) \,.
\end{equation}
The above equation allows us to obtain analytical expressions for stationary quantities, such as mean square displacements (Bfactors) and
other observables, including correlations, response functions, and entropy production, the latter quantifying the distance from 
equilibrium in a physical or biological system. In addition, our model allows us to explicitly evaluate the transfer entropy, 
which measures causal relationships among protein regions.
The first step to validate the model is verifying that the aGNM represents a non-equilibrium stationary system, as indicated by a non-vanishing entropy production rate.

\subsection{Entropy production
\label{sec:entropy_prod}}
Given that the PDZ2 domain is influenced by an active bath, which maintains the molecule in a fluctuating, out-of-equilibrium state, it is natural to examine the entropy production. 
As customary, this quantity measures the entropy generated by irreversible processes and serves to identify the key parameters responsible for driving the system away from equilibrium.

For our aGNM approach to PDZ2 fluctuations, the entropy production rate (EPR) is given by the product of active noises and 
residue displacements (see \ref{app:EPR})
$$
\dot S= \frac{mg}{T_0} \sum_{ij} \mathbb{K}_{ij}  \left\langle w_i(t)   x_j(t) \right\rangle 
$$
As shown in \ref{app:EPR} and \ref{app:correla}, the solutions for the normal modes, $Q_k(t)$ and $P_q(t)$, 
allow us to obtain the following explicit formula for the entropy production:
\begin{equation}
\dot S= k_B \frac{T_a }{T_0} g \sum_{k} \dfrac{\tau_a}{1+ \mu_k \tau_a}  \sum_{ij} u_i(k)\, \mathbb{K}_{ij} u_j(k)
\label{eq:sotS}
\end{equation}
and finally
$$
\dot S= 3k_B \gamma\,\frac{T_a }{T_0}  \sum_{k} \dfrac{\mu_k  \tau_a}{1+ \mu_k \tau_a}.  
$$
The factor $3$ stems from the well-known degeneracy due to the isotropy of the GNM. 
As expected for an effective description of a protein in a non-equilibrium environment, the entropy production is positive. 

The rate $\dot{S}$ depends linearly on the ratio $T_a/T_0$, active-bath over the thermal-bath temperature, as well as the active time scale $\tau_a$ consistently with previous studies on entropy production calculated in active crystals, i.e. ordered networks with a first neighboring connectivity~\cite{caprini2023entropons}. 
The factor $T_a/T_0$ indicates that EPR increases with higher active temperature $T_a$, as expected, because a stronger active bath drives the 
system farther from equilibrium, leading to greater dissipation. 
The memory time scale, $\tau_a$, appears both in the numerator and the denominator of the summation terms, 
coupled to the mode relaxation timescales $\lambda_k^{-1}$.  
This suggests that as $\tau_a$ vanishes, the entropy production also vanishes since the active bath becomes an equilibrium heat bath. 
In summary, $T_a$ controls the strength of active fluctuations, while $\tau_a$ controls their persistence,  
and their interplay determines the extent to which the system remains out of equilibrium, with a balance between rapid dissipation 
(small $\tau_a$) and more sustained non-equilibrium effects (large $T_a$).

\subsection{Correlation functions}
We are interested in the correlated displacements of PDZ2 residues 
for two reasons: i) the correlations identify concerted or coordinated fluctuations among 
protein regions that are a first step to assess the validity of relationship~\eqref{eq:StrDynFunk};
ii) correlations fully characterize the statistics of Gaussian processes, as 
they are the fundamental building blocks for constructing all statistical observables.
In our aGNM, these are given by  
\begin{equation}
\langle x_i(t)x_j(t')\rangle = 
\sum_{k,p} \langle Q_k(t) Q_p(t') \rangle\;u_i(k)\,u_j(p)
\label{eq:corr}
\end{equation}
which can be conveniently split into two components to clarify the computation, 
the active (indicated by the subscript $A$) involving $P_k(t)$ and the thermal one (subscript $T$) 
involving $\phi_k(t)$, see Eq.\eqref{eq:solQk},  
$$
\langle Q_k(t) Q_p(t') \rangle = \langle Q_k(t) Q_p(t') \rangle_A + 
\langle Q_k(t) Q_p(t') \rangle_T.
$$
This is possible since active and thermal noises are considered independent.
Here report the result, referring to~\ref{app:correla} for the explicit derivation.
The active contribution reads
\begin{equation}
\langle x_i(t)x_j(t')\rangle_A = 
\dfrac{3 k_B T_a}{m} \sum_{k}\; u_i(k) u_j(k) 
\frac{\tau_a^2}{\mu_k\tau_a (\mu_k^2\tau_a^2 - 1)} \left(\mu_k\tau_a e^{-|t-t'|/\tau_a} -e^{-\mu_k |t-t'|}\right) \,.
\label{eq:corrA_time}
\end{equation}
The thermal contribution has the following expression
\begin{equation}
\langle x_i(t)x_j(t')\rangle_T =
\dfrac{ 3 k_B T_0}{m \gamma} \sum_k u_i(k) u_j(k) \dfrac{e^{-\mu_k |t-t'|}}{\mu_k}\;.
\label{eq:corrT_time}
\end{equation}
Again, the factor $3$ is due to the degeneracy of the GNM, which identifies the $x,y,z$ directions and, for this reason, 
can be considered a sort of topological approach to protein structures.
From Eq.~\eqref{eq:corrA_time}, we can deduce that activity slows down the relaxation of the system; in fact, 
if $\tau_a$ is large, then $\mu_k \tau_a \gg 1$ and the second exponential term dominates the first one. 
 
Since the stationary state implies that correlations depend only on the time difference, $|t-t'|$, in the following, 
we can safely focus on the case $t'=0$, for notation simplicity.
It is convenient to rewrite the full expression of the correlation in a more compact form 
\begin{equation}
C_{ij}(t) = \langle x_i(t)x_j(0)\rangle = 
\dfrac{ 3 k_B T_0}{m\gamma} \sum_k u_i(k) G(k,t) u_j(k), 
\label{eq:corrGk}
\end{equation}
by defining
\begin{equation}
G(k,t) = \dfrac{e^{-\mu_k t}}{\mu_k} + \dfrac{T_a}{T_0} \frac{\gamma\tau_a^2}{\mu_k\tau_a (\mu_k^2\tau_a^2-1)} \left(\mu_k\tau_a e^{-t/\tau_a} -
e^{-\mu_k t}\right) \,.
\label{eq:Gk}
\end{equation}
Equal time correlation $\langle x_i(0) x_j(0) \rangle$ is directly related to the fluctuations of the 
PDZ2, 
\begin{equation}
   C_{ij}(0) = C^T_{ij}(0) +  \dfrac{3 k_B T_a}{m g} \sum_k \dfrac{u_i(k)\,u_j(k)} {\lambda_k} 
\;\dfrac{\gamma\tau_a}{1 + g \lambda_k \tau_a/\gamma}
\label{eq:deltaC}
\end{equation}
where $C^T_{ij}(0)$ denotes the pure thermal correlation: Eq.\eqref{eq:corrT_time} evaluated at $t=t'$. 
This form clearly shows how activity corrects the thermal correlations with a term modulated by the factor $T_a$ 
and depending on both $\tau_a$ and $\lambda_k$.

\section{Validation of the Active Gaussian Network Model
\label{sec:valida}}
In this section, the aGNM for PDZ2 is validated against experimental data by computing the predicted 
B-factors and comparing them with experimental B-factors reported in the 3LNX PDB-file.

To understand the impact of the activity in the fluctuations of PDZ molecule, it is natural to start from the computation of 
mean-square-diplacement (theoretical B-factors), obtained as the diagonal elements of the equal time 
correlation matrix, $C_{ii}(0) = \langle \mathbf{r}^2_i(t) \rangle$  
\begin{equation}    
\mathrm{BF}_{i} = \langle \mathbf{r}^2_i(t) \rangle = 
C^T_{ii}(0) +  \dfrac{3 k_B T_a}{m g} \sum_k \dfrac{u_i^2(k)} {\lambda_k} 
\;\dfrac{\gamma\tau_a}{1 + g \lambda_k \tau_a/\gamma}\,.
\label{eq:BFact}
\end{equation}
The correction to the thermal B-factors, $C^T_{ii}(0)$, due to the activity is proportional to $T_a$, and the thermal behavior 
is recovered when $\tau_a\to 0$ or $T_a\to 0$. 
This expression represents a generalization for the position fluctuations obtained for an active crystal~\cite{caprini2023entropons} 
to a network with a more complex connectivity.

A technical advantage of the aGNM over the standard GNM relies on its greater variability, due to the three parameters $g, T_a, \tau_a$ that can be adjusted to capture the experimental B-factors more accurately.
The parameter $g$, as in standard GNM, sets the energy scale and influences the amplitude of fluctuations; 
$T_a$ can be varied to tune the weight of non-thermal contributions.
Lastly, $\tau_a$ represents an extra correlation time controlling the memory of residue displacements. 
By refining these parameters, the model can be calibrated to reproduce the dynamic behavior observed in experiments.
Since this study focuses on qualitative aspects, we do not adjust the parameters of the aGNM to match experimental B-factors, rather, 
we apply a simple rescaling with the mean    
$$
\frac{\mathrm{BF}_i}{\sum_j \mathrm{BF}_j / N} \,.
$$  
This normalization automatically aligns the curves, facilitating direct comparison between predicted and experimental B-factors.

\begin{figure}[t!]
\centering
\includegraphics[width=\textwidth]{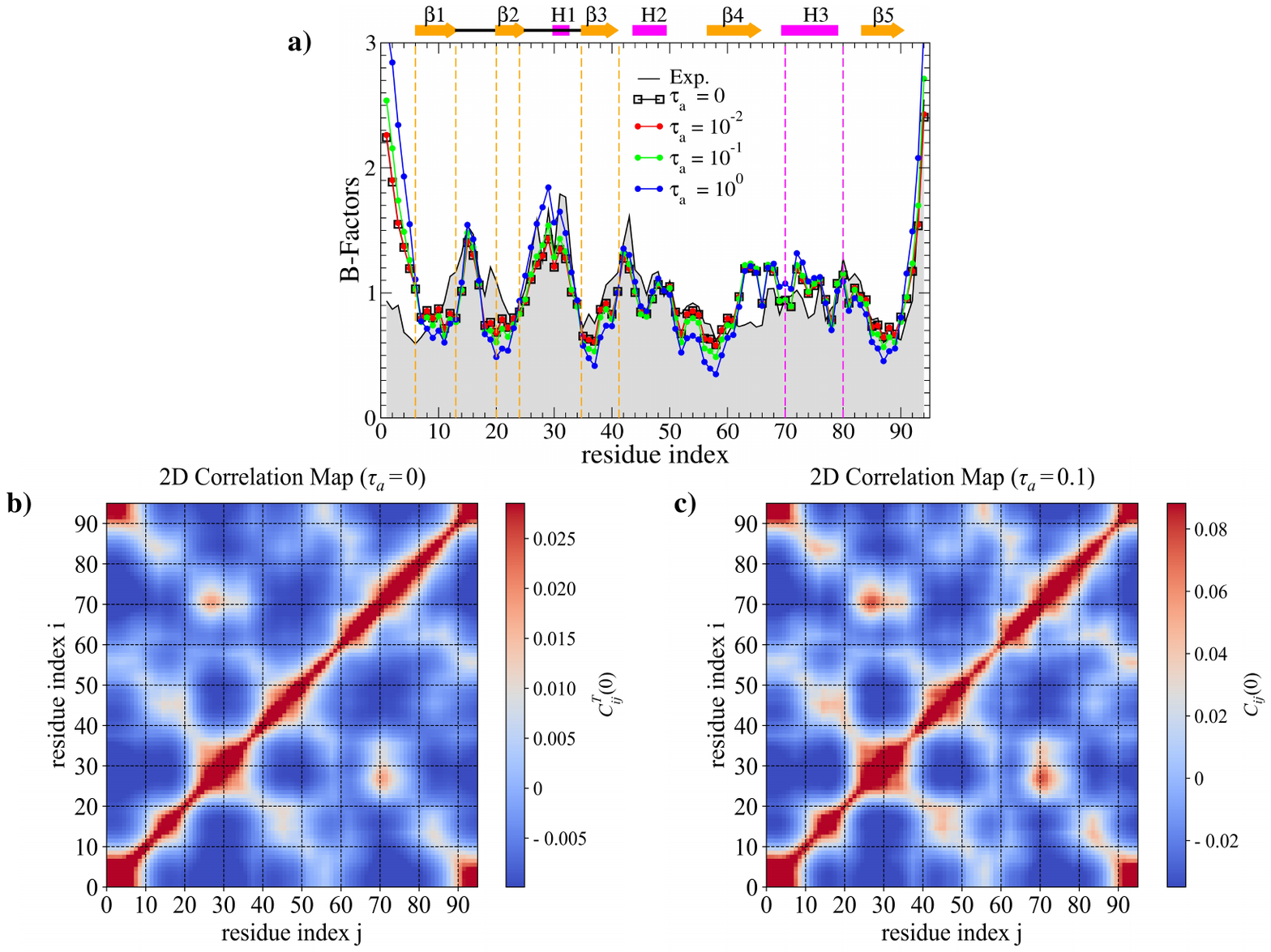}
\caption{Performance check of aGNM.
Panel (a) comparison of: i) thermal B-factors, obtained at $\tau_a=0$; 
ii) nonequilibrium Bfactors, Eq.\eqref{eq:BFact}, for values $\tau_a=10^{-2},10^{-1}, 1.0$, $T_a=20$, $T_0=1$, 
and iii) experimental BFactors (from PBDid:3LNX), shaded profile. 
To achieve the collapse, all data are rescaled with the mean $\mathrm{BF}_i/\sum_j \mathrm{BF}_j/N$.
The PDZ2 secondary-structure elements are reported on top of the panel for identification.
Color map of thermal correlations b) and active correlation c).    
\label{fig:agnm_test} 
}
\end{figure}

To simplify the analysis, we vary only $\tau_a$ in the B-factor comparison, as $T_a$ results in a simple amplification.
Figure \ref{fig:agnm_test}(a) displays the profile derived by Eq.\eqref{eq:BFact}, for different $\tau_a$,  
clearly indicating that the aGNM retains the essential characteristics of the B-factor profile observed in the thermal GNM. 
It reasonably captures the intrinsic, protein-like dynamics of the PDZ2 domain, thereby avoiding the introduction of any spurious or 
artificial patterns due to arbitrary coupling of modes. 
Therefore, the aGNM offers the significant advantage to magnify the peak-valley patterns, 
particularly in regions associated with structural flexibility.
This result can be explicitly derived by expanding Eq.\eqref{eq:deltaC} in a Taylor series for small $\tau_a$, yielding
\begin{equation}
\begin{aligned}
   C_{ij}(0) &= C^T_{ij}(0) +  \dfrac{3 k_B T_a}{m g} \sum_k \dfrac{u_i(k)\,u_j(k)} {\lambda_k} \sum_{n=0}^{\infty}(-1)^n\gamma\tau_a \Big ( \dfrac{g\lambda_k\tau_a} {\gamma}\Big)^n \\
   &=C^T_{ij}(0)\Big(1+\dfrac{T_a}{T_0}\gamma\tau_a \Big) - \dfrac{3 k_B T_a}{m g} \tau_a^2\delta_{ij}
   -\dfrac{3 k_B T_a}{m g} \tau_a^2 \sum_{m=1}^{\infty}(-1)^m\Big(\dfrac{g\tau_a}{\gamma}\Big)^m [\mathbb{K}^m]_{ij}.
\label{eq:C_series}
\end{aligned}
\end{equation}
As a result, in the small persistence time, $\tau_a$, the model primarily reflects equilibrium behavior, where fluctuations are largely thermal, except for the amplification depending on $T_a$ and $\tau_a$.

The red regions in the correlation color maps of Fig.\ref{fig:agnm_test} (b) and (c), indicating positive correlations, localize around the points of the contact map; this suggests that residues with direct (native) interactions tend to exhibit positive correlations. 
This occurs because mechanical and structural constraints enforce cooperative motion between these residues.
By contrast, as $\tau_a$ increases, the model approaches a non-equilibrium (active) regime, where fluctuations grow in amplitude, especially in more flexible or ``floppy'' protein regions.
In conclusion, these non-equilibrium dynamics accentuate protein flexibility without disrupting the fundamental protein-like patterns of 
the system.

\subsection{Equal-time correlation analysis
\label{sec:correla}
}
\begin{figure}[t!]
\centering
\includegraphics[width=15.0cm]{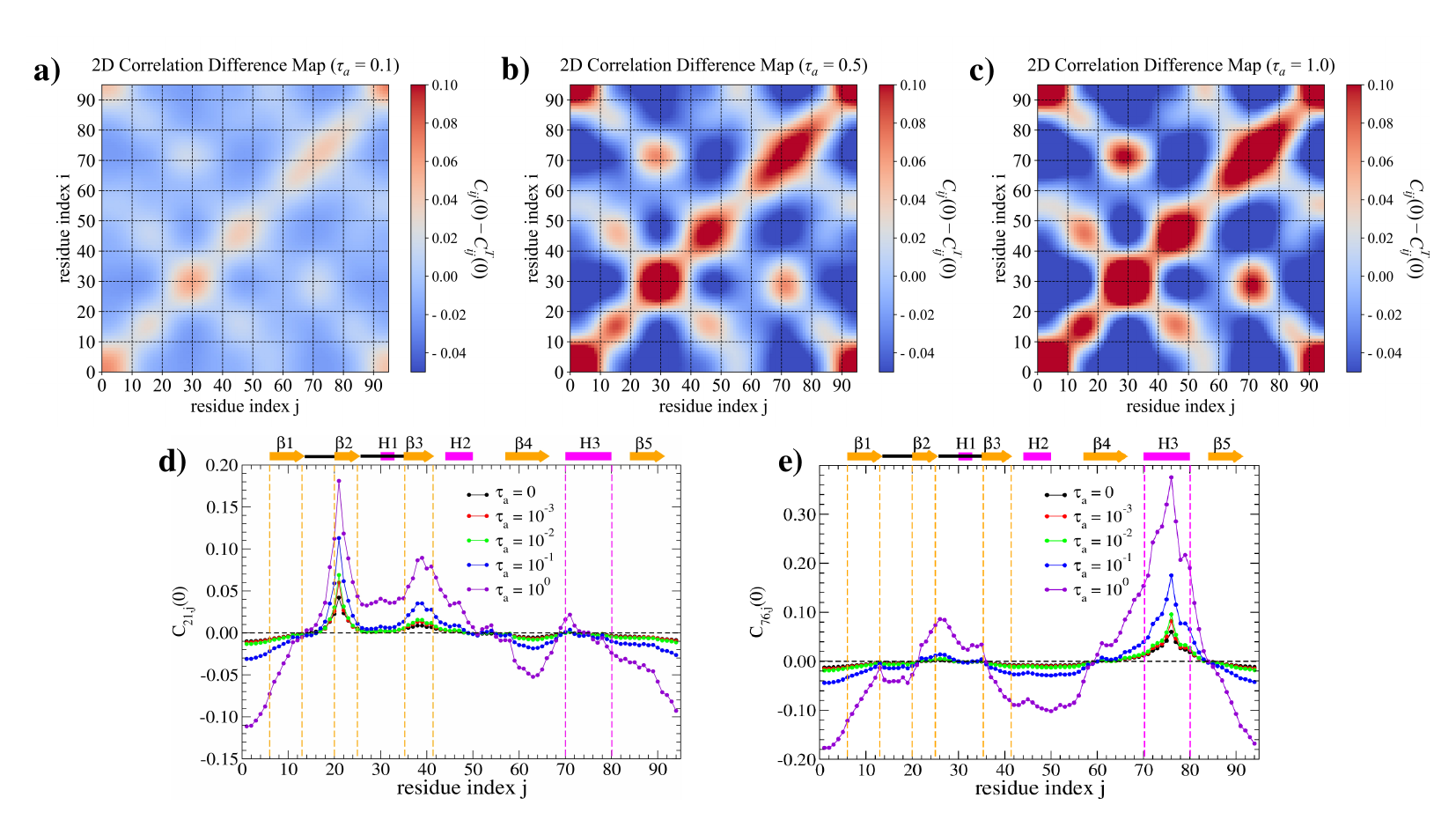}
\caption{Color maps of the difference, $C^A_{ij}(0)-C^T_{ij}(0)$, 
between active and thermal correlation, Eq.\eqref{eq:deltaC}, between PDZ aminoacids, 
for three activities: $\tau=0.1$ (a), $\tau=0.5$ (b) and $\tau=1.0$ (c). 
Data are obtained at $T_a=20\,T_0$ with $T_0=1$.  
Correlation profiles $C_{i=21,j}(0)$, panel d), and $C_{i=76,j}(0)$, panel e) measuring how site 21 and 76 establish correlation 
with the rest of aminoacids $j=1,\ldots,94$. 
The curves are obtained for different $\tau$ spanning three orders of magnitude (legend), 
with $\tau=0$ corresponding to the thermal fluctuations of PDZ2.
\label{fig:agnm_compare} 
}
\end{figure}

In this section, we analyze the correlation matrix of PDZ2 residue displacements, Eq.\eqref{eq:deltaC}, to better assess the impact of activity on protein dynamics. 
In particular, we compare the correlation aGNM with the correlation computed with a purely thermal GNM, as shown in Fig.~\ref{fig:agnm_compare}.
Equal-time correlation analysis can reveal residue pairs that ``co-move'' and is essential in understanding the concerted or coordinated motion of protein regions at the same time, suggesting how residues may organize or participate in collective behavior for biological functions \cite{Karpluscollective}. 
In particular, correlated movements can identify networks of residues that transmit signals or motions across the structure, revealing potential allosteric pathways and communication networks within the protein \cite{hub2009detection}.

To assess how aGNM corrects the thermal GNM, we present in panels (a), (b), and (c) of Fig.~\ref{fig:agnm_compare} 
the color map of the deviation, $C_{ij}(0)-C^T_{ij}(0)$, {\em i.e.} the pure activity contribution to equal-time correlations.
These maps show that, as the activity $\tau_a$ increases, $C_{ij}(0) - C^T_{ij}(0)$ becomes larger (brighter red) and extends over wider regions, still characterized by the widespread presence of native contacts (spring links), as seen in the contact map in Fig.~\ref{fig:3LNX}~(c).
Interestingly, a more intense red region in maps (b) and (c), roughly corresponding to $22 \le i \le 35$ and $ 65 \le j \le 78$, reveals a mutual coordination between \(\beta2\) and $H$3, suggesting a cooperative dynamic within the binding groove and highlighting its role 
in allosteric regulation. 
The same happens in region $10 \le i \le 20$ and $40 \le j \le 50$ corresponding to a cooperative fluctuation of the carboxylate binding loop, $\beta$1-$\beta$2, with $H$2 helix.  

We now compute the equal-time correlation profile for the PDZ2 domain by selecting a reference residue "$i=a$" and evaluating its correlation function $C_{i=a,j}(0)$ with every residue $j = 1,\ldots, 94$ in the PDZ2 structure.
We focus on two specific residues, $a = 21$ (placed in $\beta$2) and $a = 76$ (placed in H3), which have been identified in several experimental and simulation studies \cite{real_timePS,timeResPDZ,PSw_NMR_MD,photoswMD} as reference sites for monitoring changes in the binding-groove width of the PDZ2 structure (3LNX).
Indeed, such residues are strategically positioned on opposite sides of the binding groove, perpendicular to its axis, making them perfect for capturing the dynamics of the groove deformation upon ligand binding.
This analysis reveals how fluctuations at the chosen reference site are correlated with those across the entire protein, 
thereby providing insight into the spatial distribution of correlations along the PDZ2 domain.

A first inspection of Figures \ref{fig:agnm_compare}(d) and (e) suggests that correlation profiles become more 
pronounced as $\tau_a$ increases. 
In particular, panel (d) shows that high activity enhances the correlations between residue $21$ in $\beta$2 and those in 
$\beta$2-H1-$\beta$3 loop, suggesting a strong cooperative motion in this region. While residue $21$ is anti-correlated with 
the strand $\beta$4 and tail-terminals.
Residue $76$ in panel (e) also establishes correlations with the $\beta$2-H1-$\beta$3 loop and anticorrelations with the region 
involving helix H2 and the elements $\beta$3 and $\beta$4. 
Therefore, the aGNM correlations highlight some role of $\beta$2-$H$1-$\beta$3 within the allosteric network described in section \ref{sec:pdz}.
 
We conclude this section by observing that B-factor results and equal-time correlation analysis confirm that aGNM is a reasonable extension of the thermal 
GNM: although it preserves the structural properties of the standard GNM modes, modifies their statistical properties.
This means that the spatial patterns of fluctuations remain unchanged, but their relative magnitudes and impact on molecular dynamics are different. 
As a result, aGNM offers a generalized representation of the system's fluctuations, which could be more effective in capturing PDZ2 allosteric effects compared to the conventional GNM.
A more detailed correlation analysis will be presented in the following section where dynamical impact of activity on fluctuations will be discussed.

\section{Causal approach to PDZ2 allosteric dynamics
\label{sec:causal}}
In this section, we interpret the allosteric regulation of the PDZ2 domain in the general framework of causality \cite{pearl2009causality}, 
wherein ligand binding between strand $\beta$2 and $H$3 helix acts as the ``causal event'' triggering specific 
fluctuations of the PDZ2 structure facilitating its interaction with other target molecules.
As we shall see below, this causal approach to allostery will be explored using response functions and transfer entropy. 

Unlike response functions and transfer entropy, which are genuine causal indicators, correlations only measure association and not causation.
Nonetheless, it is convenient to start from correlation analysis as it provides useful preliminary information for 
identifying coordinated residue movements that are potentially involved in allosteric behavior.

\subsection{Correlation analysis}
The time-dependent correlation profiles $C_{i=a,j}(t)$ characterize how fluctuations at a reference site $a$ evolve
with respect to fluctuations of the other sites, capturing at once the ``spatial and temporal'' patterns in the PDZ2 collective motion.

\begin{figure}[t!]
\includegraphics[width=\textwidth]{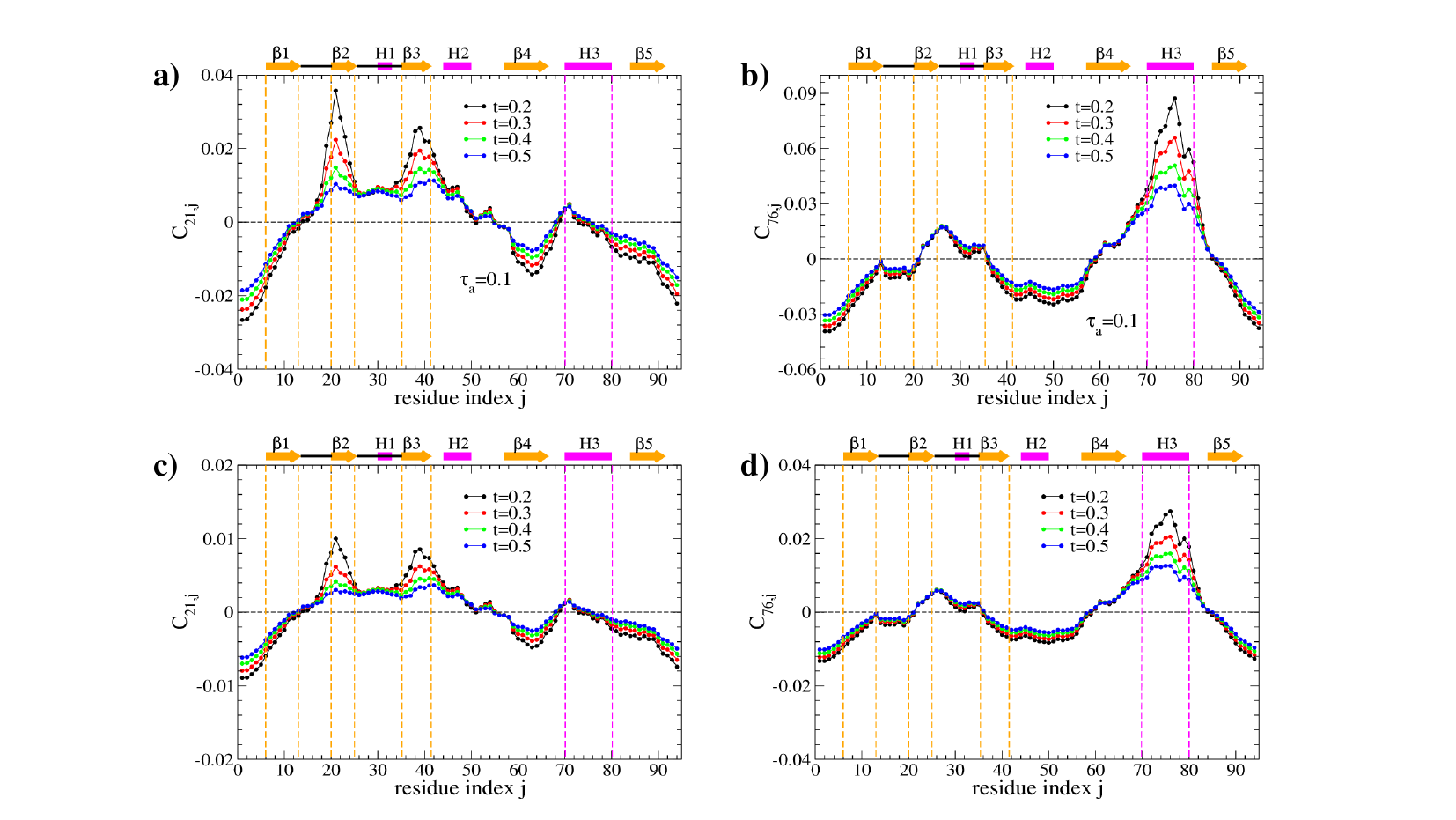} 
\caption{The correlation profiles, $C_{i=a,j}(t)$, for the aGNM model are shown in panels (a) and (b), with reference sites 
$a = 21$ and $a = 76$, respectively. 
These profiles correspond to $T_a = 20\,T_0$ ($T_0=1$) and $\tau_a = 0.1$. 
For comparison, panels (c) and (d) present the corresponding correlation profiles for the thermal GNM. 
In each panel, the profiles are sampled at the times indicated in the legend, demonstrating that their patterns 
remain consistent over a meaningful time window.
\label{fig:prof_Ct}
} 
\end{figure}

The profiles, $C_{i=a,j}(t)$, for the aGNM model, with $T_a = 20\,T_0$ and $\tau_a=0.1$,
are plotted in panels (a) and (b) of Fig.\ref{fig:prof_Ct}, corresponding to the reference sites $a = 21$ and $a = 76$ of PDZ2, respectively.
By sampling the profiles at different times, we can assess the robustness of these patterns as the correlation 
structure remains relatively unchanged over a suitable time window.
At short timescales, of course, the scenario remains similar to that observed for static correlations 
$C_{ij}(0)$ reported in panels (d) and (e) of Fig.\ref{fig:agnm_compare}.
As time progresses, the main peaks gradually decay; however, the overall pattern remains preserved due to the memory effects introduced by the activity.
For completeness and comparison, in Fig.\ref{fig:prof_Ct}~(c) and~(d), we also present the corresponding thermal profiles that appear
less resolved in structure.  

\subsection{Response analysis}
We compute the response dynamics of PDZ2 aminoacids to understand whether the normal mode decomposition can elucidate its 
allosteric regulation and correct the correlation analyses reported in Section~\ref{sec:correla}. 
Indeed, through the response functions, one can infer true causal relationships not always detected by simple correlations.

In our PDZ2 analysis, we use the response to quantify how a perturbation at the binding groove, simulating the stress induced by the ligand binding, propagates throughout the protein structure. 
The response induced on the average evolution of $x_j(t)$ by the infinitesimal initial perturbation $\delta x_i(0)$ 
is defined as
$$
R_{ij}(t) =
\dfrac{ \delta \langle x_j(t) \rangle }{\delta x_i(0)}, 
$$
where $\delta \langle x_j(t)\rangle$ is the difference between the average of perturbed and unperturbed trajectories taken over the unperturbed 
stationary distribution \cite{marconi2008fluctuation}.
For linear stochastic systems like the aGNM, the response functions between residues can be obtained from the knowledge of correlations via the simplified formula \cite{baldovin20understanding,ceccoPhysBio}
$$
{\mathbb R}(t) = {\mathbb C}(t) {\mathbb C}^{+}(0)
$$
with $\mathbb{C}(t)$ being the time-dependent covariance matrix restricted to residues only, $\mathbb{C}(t) = \langle{\mathbf x}(t){\mathbf x}^{\top}(0) \rangle$, and  ${\mathbb C}^{+}(0)$ is its pseudo-inverse at time, $t=0$.
It is important to note that this response is restricted to the residue displacements $\mathbf{x}(t)$, excluding the active-noise variables $\mathbf{w}(t)$ introduced in the aGNM dynamics~\eqref{eq:evol2}. 

\begin{figure*}[t!]
\centering
\includegraphics[width=0.7\textwidth]{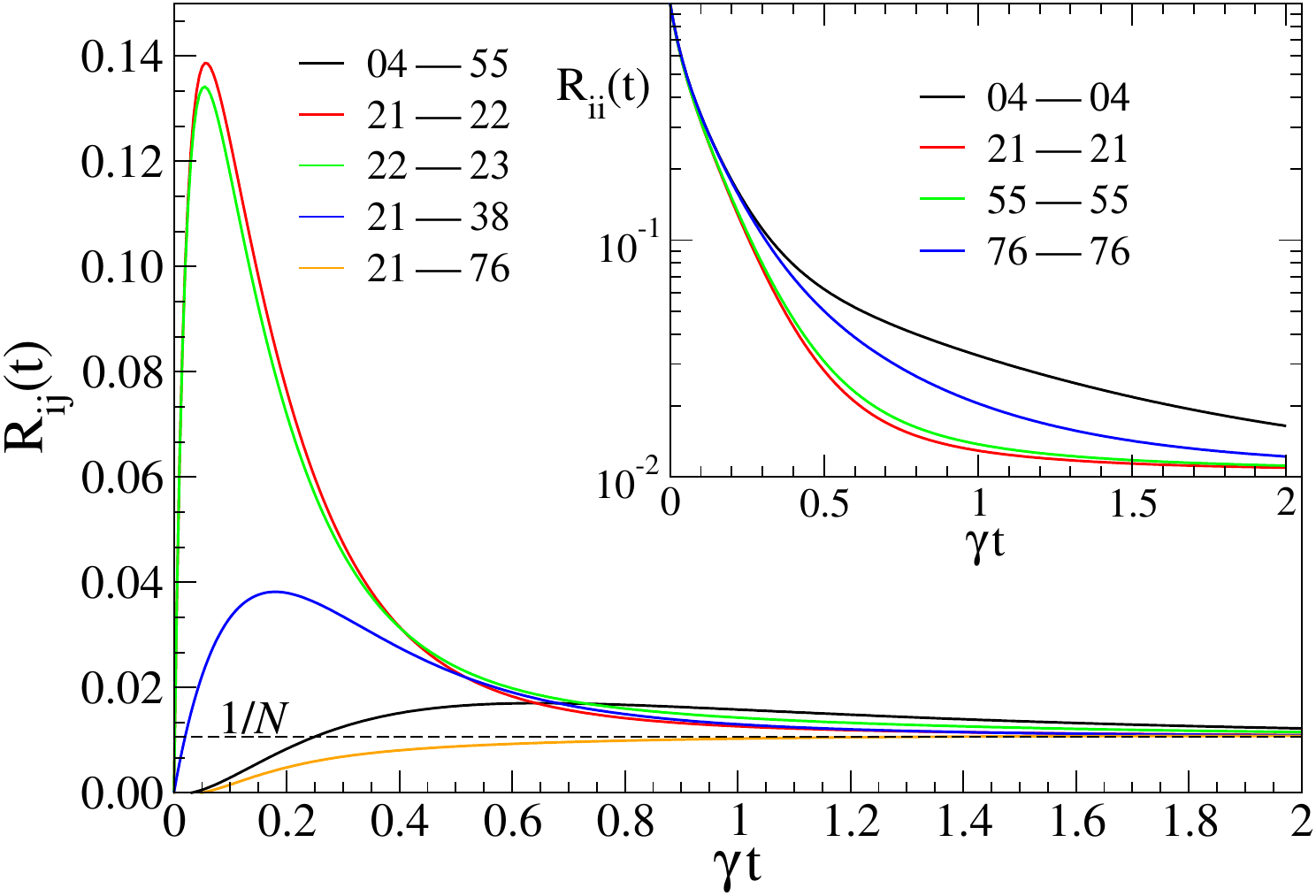}
\caption{
Time courses of responses $R_{ij}(t)$ between residues: 
$4$-$55$, $21$-$22$, $22$-$23$, $21$-$38$ and $21$-$76$. 
There are several behaviors of the response depending on the length of the path of springs (links) connecting two residues 
response curves associated with native contacts (i.e.\ in direct harmonic links), like $21$-$38$, display a peaked structure that decays to a characteristic value of $1/N$. 
Whereas, pairs of residues that are not in contact, but connected by a short path ($21$-$38$) of links, are characterized by a response that gets a lower peak and later.
Finally, pairs of residues connected by a longer path ($21$-$76$) increase monotonically from $0$ to $1/N$, they can be considered inactive in terms of response analysis.
The inset shows the non-exponential decay of the self-responses $R_{ii}(t)$ for selected residues presented in the main panel.
\label{fig:Resptime}}
\end{figure*}

Using Eq.~\eqref{eq:corrGk} and Eq.~\eqref{eq:Gk} provides the explicit formula for the response functions between each couple of PDZ aminoacids
\begin{equation}
R_{ij}(t) = \sum_{k} u_i(k)\;\frac{G(k,t)}{G(k,0)}\; u_j(k)\;.
\label{eq:Respo}
\end{equation}
Figure~\ref{fig:Resptime} reports response curves in time for some pairs od PDZ2 residues obtained by implementing numerically 
Eq.~\eqref{eq:Respo}.
The time behavior of $R_{ij}(t)$ critically depends on the length of the spring pathway connecting residues $i$ and $j$. 
Maximal responses are observed between consecutive backbone residues, such as $i$-$(i+1)$, {\em e.g.}, $21$-$22$ and $22$-$23$. 
For residue pairs in direct native contact, like $21$-$38$, the response function exhibits a pronounced peak at short times, 
followed by a relaxation toward the baseline value of $1/N$.

In contrast, residue pairs, such as $4$-$55$, that are not directly linked but are connected by a relatively short chain of springs, display a response with a lower, delayed peak. 
Finally, pairs connected by a long chain of links ($21$-$76$) show a monotonically increasing response from $0$ to $1/N$, indicating a negligible response activity.
The inset of Fig.~\ref{fig:Resptime} illustrates the decay of self-responses, $R_{ii}(t)$, for selected residues shown in the main panel, highlighting the non-exponential behavior induced by memory effects from the active bath.

\begin{figure}[t!]
\centering
\includegraphics[width=\textwidth]{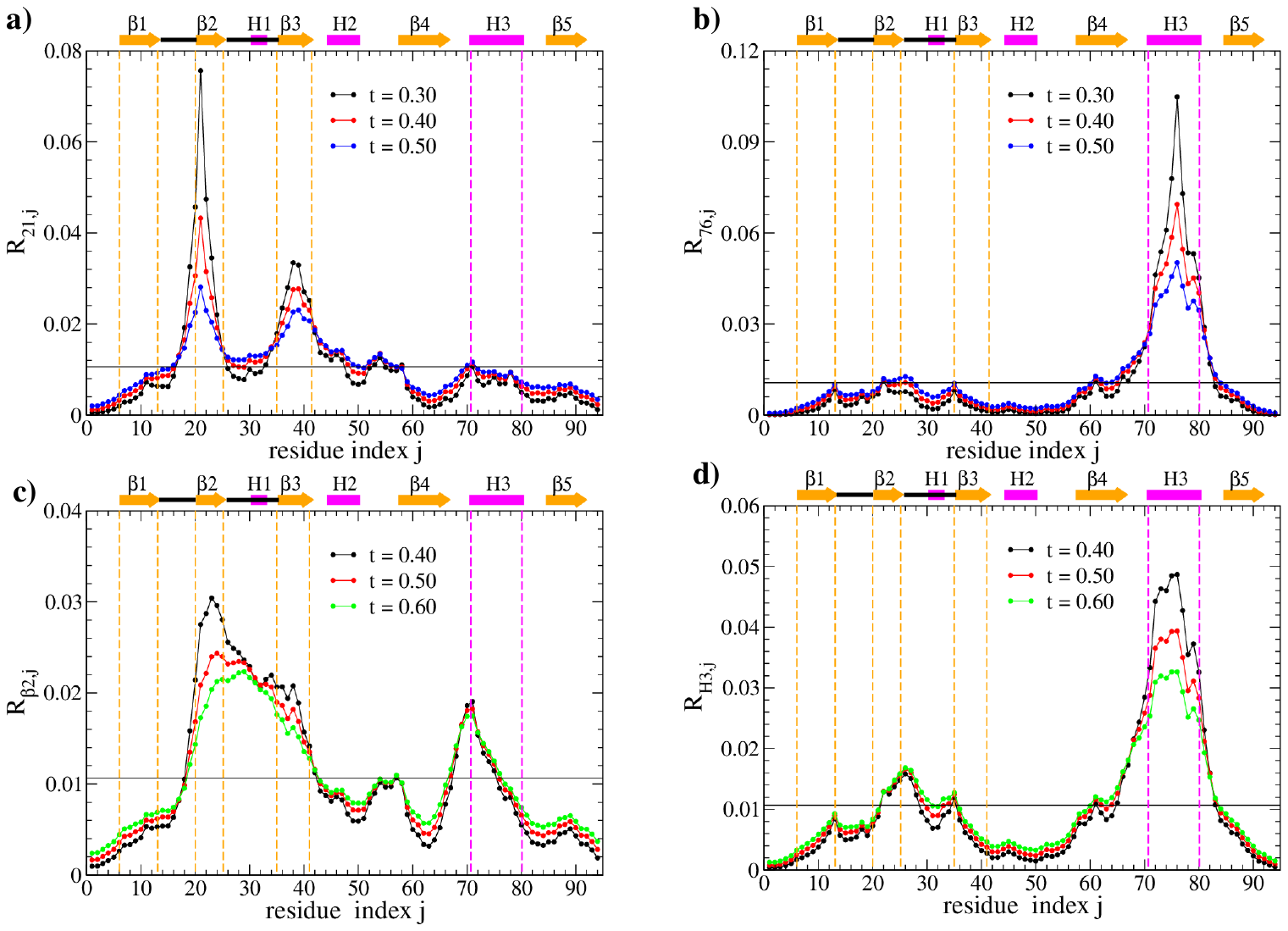}
\caption{Response profiles $R_{i=a,j}$ for $a=21$ panel (a) and $a=76$ panel (b) at different sampling times, listed in the legend.
Paned c) and d) report the block-response profile in Eq.\eqref{eq:Rblock} for the blocks $\beta$2 and $H$3 respectively, 
both flanking the binding groove.   
Data are obtained with an aGNM with $\tau_a=0.1$ and 
$T_a = 20\,T_0$.
\label{fig:prof_Rt}
}
\end{figure}

By selecting the same reference residues $a=21$ and $a=76$ in the binding groove, we derive response profiles $R_{i=a,j}(t)$ that describe how perturbations at these sites (i.e., from the binding groove) are transmitted throughout the PDZ2 structure.
Unlike correlation profiles, the analysis of these profiles allows us to exclude regions that are not in a true causal relationship, thereby selecting only residues involved in the allosteric behavior exhibited by PDZ2.
Panels (a) and (b) of Fig.~\ref{fig:prof_Rt} plot the response profiles for $a=21$ and $a=76$, respectively.
Panel (a) shows that a perturbation at site $21$ rapidly propagates to neighboring residues in the $\beta$2 strand, as seen 
by the sharp and spreading peak around $21$. 
The peak on $\beta$3 reveals that the perturbation at $21$ also affects residues in that strand.
In contrast, the loop $\beta$2-H1-$\beta$3, belonging to the allosteric network, exhibits a modest and retarded response.
Finally, the start of helix H3 is slightly affected by the perturbation in 21, suggesting minimal evidence of binding groove deformation.
The response profile $R_{a=76,j}$ in panel (b) reveals that a perturbation at residue $76$ rapidly extends over the H3 helix, as seen by the spreading of the sharp peak in H3. 
Remarkably, the impact of residue $76$ on other regions of PDZ2 remains negligible, as its profile falls below the $1/N$  baseline, except for a minor effect on the $\beta$2 strand, again suggesting a deformation of the binding groove.

These observations seem to indicate that helix H3 plays a rather limited role compared to $\beta$2 in mediating allosteric communication within PDZ2.
To support this claim, it is essential to go beyond the responses generated by pointwise perturbations and instead examine the collective response from the whole  $\beta$2 and H3 regions. 
We thus define the block response as the mean of response profiles $R_{ij}$ over a generic secondary structure $X$ of interest 
\begin{equation}
R_{X,j} = \dfrac{1}{{\cal N}_X} \sum_{i\in X}  R_{ij}\,, 
\label{eq:Rblock}
\end{equation}
where ${\cal N}_X$ denotes the number of residues contained in $X$.
This procedure yields a collective response that characterizes the overall dynamical impact of the structure $X$, rather than focusing on the response at individual sites.

The profiles, $R_{X,j}$, help in comparing how $\beta$2 differs from helix H3 in terms of their influence on the rest of the PDZ2; see Fig.\ref{fig:prof_Rt}.
Panel (c) refers to $\beta$2, the first edge of the binding groove, and shows that the collective perturbation of this strand mainly extends on the loop $\beta$2-H1-$\beta$3 and also impacts the region across $\beta$4 and helix H3.
Therefore, one can deduce that a deformation of the binding groove due to $\beta$2 mainly affects the loop $\beta$2-H1-$\beta$3 and partially the opposite side of the binding groove, H3.
This result is different from the analysis of Fig.~\ref{fig:prof_Rt}~(a) obtained by the single-site response that cannot represent the behavior of an entire secondary structure element.
Finally, Fig.~\ref{fig:prof_Rt}~(d) shows that the response to the perturbation in $X =$ H3 is primarily observed in the loop structure $\beta$2-H1-$\beta$3, 
underscoring the sensitivity of this loop.
Interestingly, the finding that the $\beta$2-H1-$\beta$3 loop responds to perturbations in both $\beta$2 and H3 suggests that its allosteric role is driven by the opening of the binding groove, in agreement with Refs.~\cite{kozlov2002solution,ivarsson2012plasticity}.

In the next section, we complement response analysis with transfer entropy results to refine our proposal of interpreting PDZ2 allostery in the framework of causality.

\subsection{Transfer-Entropy analysis}
Transfer entropy (TE), introduced by Schreiber in Ref.\cite{schreiberTE2000} in the context of dynamical system and in information theory by Palus \cite{palusTE2001}, is a causal indicator quantifying how much the past behavior of a variable $X$ helps to predict the future behavior of another variable $Y$ beyond what can be predicted by only observing the past of $Y$. 
In other words, TE characterizes how information flows from
the time evolution of $X(t)$ to the evolution of $Y(t)$ 
indicating a directional influence.
This concept is useful for analyzing protein dynamics \cite{erman2017causalGNM,erman2017causalMD,ceccoPhysBio}, 
as it provides directional influence between different regions or residues in a protein molecule.

In our study, TE analysis complements the correlation and response results, as it might detect causal and complex 
interdependencies among PDZ2 residues during their organized fluctuations, which is crucial information to explain PDZ2 allosteric behavior.
Moreover, the non-reciprocal nature of TE provides a directional view of information flow in PDZ2 dynamics, distinguishing ``driving'' from ``driven'' residues.
In this classification, a driving residue or region is generally a donor of entropy, meaning that its motion determines an information flow toward other residues (regions) that are instead considered driven or acceptors. 
The concept can also be reformulated in terms of uncertainty because knowledge of the dynamical state of residue, ``$j$'', can reduce uncertainty about the dynamical state of residue ``$i$''.

According to the definition given in Ref.\cite{schreiberTE2000}, the explicit TE expression from residue $j\to i$, is the quantity
\begin{equation}
\mathrm{TE}_{j\to i}(t) = 
\bigg\langle 
\log 
\frac{P[x_i(t)\,|\,x_i(0),x_j(0)]}
{P[x_i(t)\,|\,x_i(0)]}
\bigg\rangle \,,
\label{eq:defTRE}
\end{equation}
where the angular brackets indicate the average over the joint probability 
density $P[x_i(t),x_i(0),x_j(0)]$, while and $P[x_i(t)\,|\,x_i(0),x_j(0)]$ and
$P[x_i(t)\,|\,x_i(0)]$ are the probability
densities of $x_{i}(t)$ conditioned to the previous values $x_i(0)$ and/or $x_j(0)$.
Due to the 
presence of conditional probabilities, TE is not symmetric, $\mathrm{TE}_{j\to i} \neq \mathrm{TE}_{i\to j}$: the conditioning and conditioned events cannot be exchanged 
because they do not play equivalent roles.
Morever, TE$_{j\to i}$ identically vanishes for $i=j$ as  
$P[x_i(t)\,|\, x_i(0),x_j(0)] =  P[x_i(t)\,|\, x_i(0)]$.

For Gaussian systems like both aGNM and GNM, there is a simple way to evaluate the TE among residues by using the correlations \cite{sun2015causal,erman2017causalGNM,ceccoPhysBio}, Eq.\eqref{eq:corr},
\begin{equation}
\mathrm{TE}_{j\to i}(t) =  -\frac{1}{2} \ln\bigg(1 - \frac{\alpha_{ij}(t)}
	{\beta_{ij}(t)}\bigg)
	\label{eq:TRE}
\end{equation}
where  
\begin{equation}
	\alpha_{ij}(t) = [C_{ii}(0) C_{ij}(t) - C_{ij}(0) C_{ii}(t)]^2,
\end{equation}	
and 
\begin{equation}
\beta_{ij}(t) = [C_{ii}(0)C_{jj}(0) - C^2_{ij}(0)] 
	            [C^2_{ii}(0) - C^2_{ii}(t)]\,.
\end{equation}
A sketch of the derivation is contained in Ref.~\cite{sun2015causal,ceccoPhysBio} and  
formula~\eqref{eq:TRE} is a more compact notation of the expression reported in Ref.~\cite{erman2017causalMD}.

The asymmetry $\alpha_{ij}(t)\ne \alpha_{ji}(t)$ and 
$ \beta_{ij}(t) \ne \beta_{ji}(t)$ is a straightforward consequence of the TE asymmetry emerging also in the Gaussian formulation. 
In addition, it is immediate to show that $\mathrm{TE}_{j\to i}(\infty) = 0$, either by definition, Eq.~\eqref{eq:defTRE}, 
invoking the independence of events far away in time, or using the correlation decay at large times in Eq.~\eqref{eq:TRE}.
Analogously, one expects $\mathrm{TE}_{j\to i}(0) = 0$.

Like for the response analysis, we restrict our investigation of TE profiles to reference sites $a=21$, $a=76$ and to the binding groove elements 
$\beta$2 and H3.
As suggested in Ref.~\cite{erman2017causalGNM}, it is preferable to define the net entropy transfer (NET) between residue $i$ and $j$ as the difference between the entropy transfer from $i \to j$ ($i$ donor and $j$ acceptor) and the entropy transfer from $j \to i$ ($j$ donor and $i$ acceptor)    
\begin{equation}
\delta\mathrm{TE}_{i,j} = \mathrm{TE}_{i\to j} - \mathrm{TE}_{j\to i}\;. 
\label{eq:net}
\end{equation}  
If $\delta\mathrm{TE}_{i,j} > 0$, residue $i$ primarily gives entropy, meaning its fluctuations influence the future dynamics of $j$. 
In this case, $j$ acts as a {\em driver} in the system. 
Conversely, if $\delta\mathrm{TE}_{i,j} < 0$, residue $i$ is more influenced by the motion of $j$ than viceversa, indicating that $i$ is conditioned by $j$.

\begin{figure}[t!]
\centering
\includegraphics[width=\textwidth]{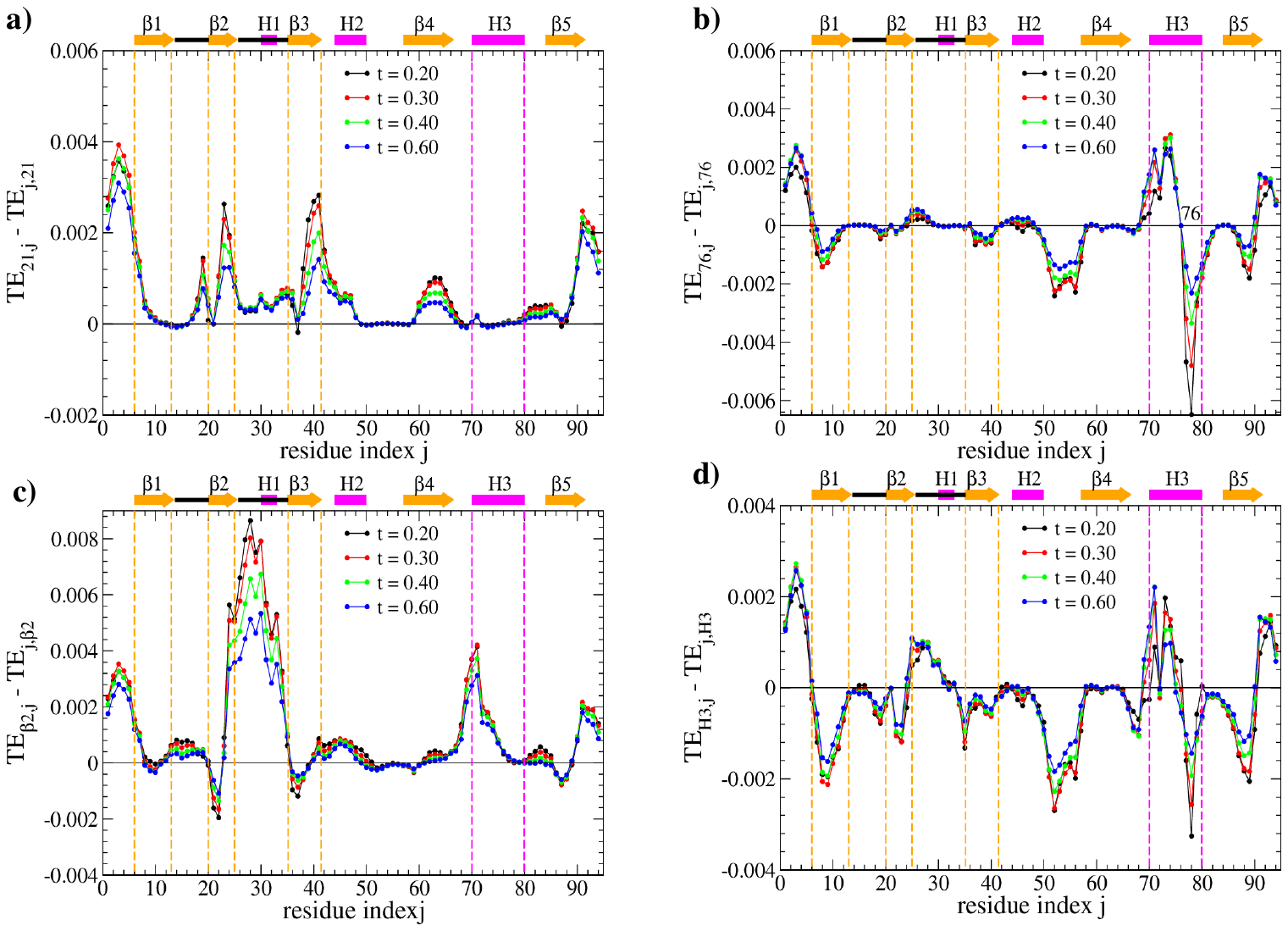}
\caption{
Net entropy transfer (NET) profiles referring to residue 21, panel (a) and 
residue 76 panel (b).
NET of residue $21$ is generally positive; thus, $21$ appears to be mainly a donor of entropy,
it drives or controls the motion of regions $\beta$3, $\beta$4, N and C-terminal tails. Moreover $21$
shows a modest impact on the loop $\beta$2-H1-$\beta$3. There is no well-defined sign of NET associated with residue $76$ in panel (b), which seems to indicate a more ambiguous role of that residue in the allosteric control of PDZ2. Panels c) and d) report block NET profile, Eq.~\eqref{eq:blockNET}, over the secondary structures $\beta$2 and H3.   
\label{fig:NET}
}  
\end{figure}

For the reference residues 21 and 76, the NET profiles $\delta\mathrm{TE}_{i=21,j}$ and $\delta\mathrm{TE}_{i=76,j}$, obtained by varying $j$, 
are presented in Fig.~\ref{fig:NET}~(a) and~(b).
In panel (a), residue $21$ seems to act like a source of information, as its NET profile to other residues $j$ is mainly positive.
Peaks in this graph identify those residues $j$ most driven by residue $i=21$, highlighting potential allosteric pathways emerging from $21$.
In particular, the residue $21$ appears to mainly control the fluctuating behavior of far structures $\beta$3 and $\beta$4 and partly of H2.
Moreover, residue $21$ affects the C-terminal and N-terminal tails, including modest control on loop $\beta$2-H1-$\beta$3.

The oscillation of NET profile 76 between positive and negative values suggests that residue $76$ act as both a donor and an acceptor of information: this residue donates to the tail regions and the preceding segment of H3 (residues $69$-$75$) while receives from the following segment of H3 (residues $77$-$82$). 
Additionally, it serves as an acceptor in interactions with $\beta$2 and $\beta$4. 
However, according to TE, the allosteric role of residue $76$ remains ambiguous as no clear pattern emerges to characterize its influence; 
apparently, it seems to act as an intermediate hub facilitating the redistribution of entropy.

In analogy to response analysis, it is also useful to complement the pointwise information provided by $\delta\mathrm{TE}_{i,j}$ with a block NET 
\begin{equation}
\delta\mathrm{TE}_{X,j} = \dfrac{1}{{\cal N}_X} \sum_{i\in X}  \delta\mathrm{TE}_{i,j}\;, 
\label{eq:blockNET}
\end{equation}
representing the straightforward generalization of Eq.~\eqref{eq:net} to a specific region $X$ of the PDZ,
and accounts for the overall driver/driven propensity of an entire secondary structure element.
This block NET analysis is summarized in Fig.~\ref{fig:NET}~(c) and~(d) that refer to the binding groove 
components ($\beta$2, H3). 
Figure~\ref{fig:NET}~(c) reports the $\delta \mathrm{TE}_{\beta2,j}$ profile measuring how $\beta2$ controls the rest of the PDZ2 residues. 
It shows that $\beta$2 acts as a driving element for the loop $\beta$2-H1-$\beta$3, meaning it significantly influences the fluctuating dynamics of this structural element. 
Additionally, $\beta$2 transfers entropy to the H3 helix, suggesting that $\beta$2 plays a crucial role in the fluctuations of the binding groove, behaving as a "leader". 
This might indicate that $\beta2$ is essential for the peptide binding process and contributes to the overall allosteric behavior of the PDZ.
Finally, panel~(d) confirms the role of H3 as a simple hub of entropy transfer already seen in panel~(b).

\section{Conclusions
\label{sec:concl}}
\begin{figure}[t!]
\centering
\includegraphics[width=12.0cm]{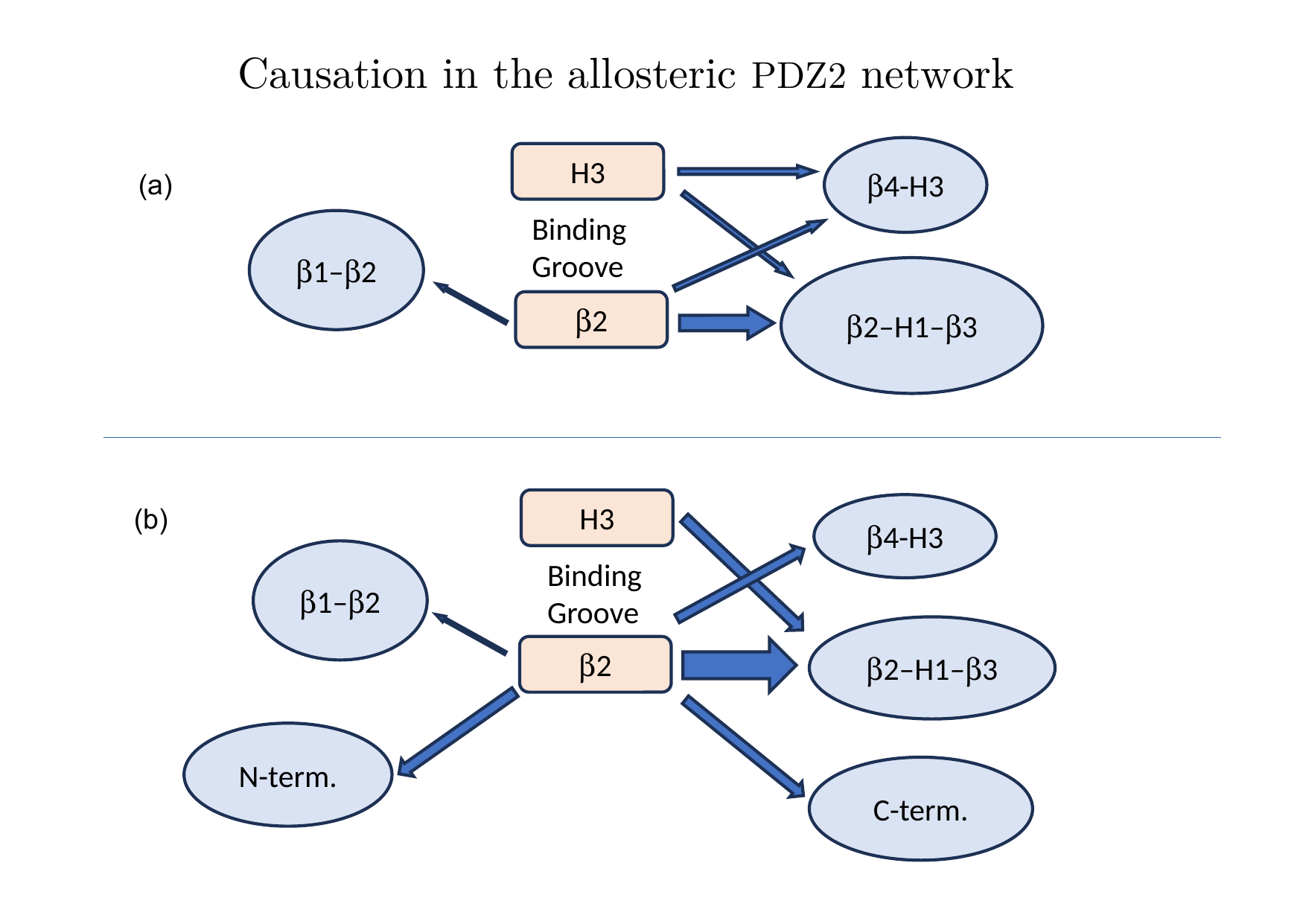}
\caption{\label{fig:netw}
Causal networks emerging from the analysis of causal indicators (a) block response (b) net entropy transfer (NET).
The arrows indicate the direction of the binding groove influence on other protein elements, and their thickness qualitatively represents the ``intensity'' of this influence as measured by the peaks in the profile of the causal indicators.
}
\end{figure}
We proposed a generalization of the Gaussian network model (GNM)
to incorporate non-equilibrium effects induced by the presence of a non-thermal 
behavior of physiological environments.
With slight flexibility in terminology, this approach has been termed active GNM (aGNM) 
by analogy to active matter systems, even though proteins are not self-propelling 
objects.

The goal of this study was to explore, within an idealized theoretical framework, whether allostery can 
emerge from non-equilibrium dynamics, as suggested in Ref.~\cite{stock2018non}. 
To test this idea, we employed the aGNM to model fluctuations in the PDZ2 domain - a well-known protein for its allosteric behavior. This approach allows us to investigate how non-equilibrium protein states might alter allosteric responses.
In this respect, we assumed that allostery itself can be framed within the {\em structure-function-dynamics} 
paradigm, ensuring that the intrinsic topology of the native fold remains the key determinant of the fluctuation 
patterns involved in PDZ2 allostery. 
 
We have tested that the aGNM reproduces experimental B-factors of PDZ2 and highlights amplified fluctuation patterns, 
emphasizing the delicate interplay between thermal and non-thermal effects in the protein dynamics.
As expected, the non-equilibrium activity retains the original structure of the GNM, {\em i.e.} the normal mode scenario, 
but alters the statistical properties of individual modes.
In this way, the relative weight of the modes can dramatically change in the allosteric communication, generally enhancing fluctuations.

In our approach to PDZ2 allostery, we adopted the hypotheses that: i) within the structure - function - dynamics paradigm, simple structure - informed models like aGNM can capture the essential allosteric dynamics; ii) allostery can be iterpreted as a causal process for which the binding of a ligand acts as the cause that triggers changes in the protein's fluctuations (effect) mediating allosteric communication.
In the PDZ2 case, the binding site controlling the protein activity is the binding-groove region flanked by 
that strand $\beta$2 and the helix H3, we consider this region as a source of the allosteric behaviour throughout the molecule.

Within the aGNM, we initially used correlation analysis to identify PDZ regions involved in allostery, then, we resorted to response function and transfer entropy analyses to identify the regions in ``true causation'' and elucidate the directional flow of information from the binding groove
to the rest of PDZ2 structure.

The response profiles reveal that perturbations at key sites - such as residues in the $\beta$2 strand - rapidly affect adjacent regions, while transfer entropy quantifies how specific residues act as dominant drivers or acceptors of dynamic information. 
These findings not only reinforce the view of proteins as complex, information-processing systems but also proposes the potential of aGNM as a computational low-cost model to be used as a lens magnifying possible allosteric pathways.
Notably, the $\beta$2-H1-$\beta$3 loop emerges as a central hub in mediating responses to perturbations, indicating its crucial role in the allosteric regulation triggered by binding groove deformations. Conversely, the relatively marginal response observed in helix H3 suggests that not all structural elements contribute equally allosteric signal propagation.

The causation scenario can be summarized by the oriented network reported in Fig.~\ref{fig:netw} 
that arises from causal-indicator analysis. In this picture, the binding groove represents the source and the arrows its influence on other PDZ2 elements. 

Our results suggest an out-of-equilibrium description of protein allostery by combining simple coarse-grained models, implementing the interplay of {\em structure-function-dynamics} with basic non-equilibrium elements.
In this context, the aGNM approach represents a simple attempt to develop an analytically tractable tool for exploring functional modes in allosteric proteins.

\section*{Acknowlegments}
F.Cecconi is very grateful to A.Vulpiani and M.Baldovin for 
enlightening discussions and valuable suggestions.
F.Cecconi, L.Caprini and G.Costantini acknowledge funding from the Italian Ministero dell'Universit\'a e della Ricerca under the programme PRIN 2022 (``re-ranking of the final lists''), number 2022KWTEB7, cup B53C24006470006.


\appendix
\section{Derivation of the entropy production formula \label{app:EPR}}

Since the PDZ domain is in contact with an active bath that drives the molecule into a fluctuating, out-of-equilibrium state, it is important to quantify this behavior by calculating the entropy production rate (EPR). The EPR measures the amount of entropy generated per unit time in a system undergoing irreversible processes. 
It serves as an indicator to quantify the distance from thermodynamic equilibrium of a system, {\em i.e.} the PDZ domain: the higher the 
EPR, the greater the irreversibility of the underlying dynamics.

To derive the expression for the total entropy production rate $\dot{S}$ within the aGNM, 
we use a path-integral method based on the following formula, where the entropy production is evaluated as the 
logarithm of the ratio between the probabilities of forward and backward trajectories~\cite{caprini2019entropy}:
\begin{equation}
 \frac{S}{k_B}  = \left\langle \log\left( \frac{\mathcal{P}_F}{\mathcal{P}_R} \right) \right\rangle \,,
\label{eq:EPformula}
\end{equation}
where $\mathcal{P}_F$ is the probability of the forward trajectory and $\mathcal{P}_r$ the probability of the time-reversed one
and the average is performed using the distribution $\mathcal{P}_F$.
To obtain an explicit formula we use the fact that the particles' trajectories are fully determined by the noise path.
The noise probability distribution has a Gaussian form for each noise governing the dynamics at each time, ${\xi}(t)$ and ${\eta}(t)$.
By using curly bracket to denote a noise trajectory from an initial time $t_0$ to a final time $t_f$ for every 
particle of the system, i.e.\ $(\boldsymbol{\xi}, \boldsymbol{\eta})$, the path probability of the noise reads
\begin{equation}
	\text{Prob}(\boldsymbol{\xi},\boldsymbol{\eta}) \propto \exp{\left( -\frac{1}{2} \sum_{i}\int_{t_0}^{t_f} dt\, \xi_i^2(t)  \right)} 
	\exp{\left( -\frac{1}{2}  \sum_{i}\int_{t_0}^{t_f} dt\, \eta_i^2(t)  \right)}\,.
\end{equation}
By expressing each noise as a function of the dynamical variables $x_i$, and $\eta_i$ by using the equations of motion~\eqref{eq:evol1} and~\eqref{eq:evol2}, one can easily switch from the probability of the noise $\text{Prob}(\boldsymbol{\xi},\boldsymbol{\eta})$ to the probability of the temporal trajectory $\mathcal{P}_F(\{\boldsymbol{x}\}, \{\boldsymbol{w}\})$ by applying a change of variables.
Since the Jacobian of this transformation does not affect the entropy production~\cite{caprini2019entropy}, we can explicitly write
\begin{eqnarray}&&
\mathcal{P}_F(\boldsymbol{x},\boldsymbol{w}) \propto \exp{\left[-\frac{m}{4 k_B T_0\gamma}\sum_i \int_{t_i}^{t_f} dt\left(\gamma \dot{x}_i + g \sum_j \mathbb{K}_{ij} x_j -\gamma w_i \right)^2\right]}
 \times \nonumber\\
&&
\times\exp{\left[ -\frac{m\tau_a}{4 k_B T_a} \sum_{i}\int_{t_0}^{t_f} dt\, \Bigl(\dot w_i+\frac{1}{\tau_a} w_i \Bigr)^2  \right]}\,.
\label{eq:app_prof_f}
\end{eqnarray}
To evaluate the time-reversed counterpart, we have to specify the behavior of the active fluctuation $w_i(t)$ under inversion of the direction of time. 
For a particle suspended in an active bath, the active fluctuation $w_i(t)$ is an random velocity and therefore should be considered odd under time reversal~\cite{shankar2018hidden}
In fact, the weight for the reverse path is found by setting $t\to -t$, $ x_i\to x_i$, $\dot x_i\to -\dot x_i$, $ w_i\to  -w_i$  and $\dot w_i\to \dot w_i$ in this expression because the time derivative is negated for the reverse path while the $x_i$ variables are not, while we assumed that $w_i$ is odd under time-reversal.
The assumption of even position and odd velocity under time-reversal transformation (TRT) appears natural. In contrast, assuming an odd active force under TRT  is less straightforward and has recently sparked debate within the active matter community~\cite{dabelow2019irreversibility, mandal2017entropy, caprini2019entropy, caprini2018comment}.
The probability of the reversed path, $\mathcal{P}_R$, reads:
\begin{eqnarray}&&
\mathcal{P}_R(\boldsymbol{x},\boldsymbol{w}) \propto \exp{\left[-\frac{m}{4 k_B T_0\gamma}\sum_i \int_{t_i}^{t_f} dt\left(-\gamma \dot{x}_i + g \sum_j \mathbb{K}_{ij} x_j +\gamma w_i \right)^2\right]}
 \times \nonumber\\
&&
\times\exp{\left[ -\frac{m\tau_a}{4 k_B T_a} \sum_{i}\int_{t_0}^{t_f} dt\, \Bigl( \dot w_i-\frac{1}{\tau_a} w_i  \Bigr)^2  \right]}\,.
\label{eq:app_prof_r}
\end{eqnarray}
By taking logarithm of the ratio of the probabilities, $ \log\left( \frac{\mathcal{P}_F}{\mathcal{P}_R} \right)$ and applying the average,  we obtain the entropy production as:
$$
\left\langle \log\left( \frac{\mathcal{P}_F}{\mathcal{P}_R} \right) \right\rangle=
  \left\langle \frac{m g}{k_B T_0}\sum_{ij} \mathbb{K}_{ij}\int_{t_i}^{t_f} \!\!dt[(w_i- \dot{x}_i)x_j]\right\rangle
  -\frac{m}{k_B T_a} \left\langle \sum_{i}\int_{t_0}^{t_f}\!\!dt\,  w_i  \dot w_i \right\rangle \,.
$$
The log ratio of the path probabilities shows that each noise source contributes a term. The total entropy production is the sum of the entropy production arising from the thermal noise $\xi_i$, governing the coordinate dynamics, and the noise $\eta_i$, controlling the active dynamics.

Finally, by dividing by the duration of the time interval, $(t_f-t_i)$ and taking the limit of an infinite interval, and discarding the boundary terms proportional to
$ \int_{t_i}^{t_f} dt \sum_{ij} \mathbb{K}_{ij} \dot{x}_i x_j $ and $ \int_{t_i}^{t_f} dt \sum_i w_i  \dot w_i$, we obtain the entropy production rate:
$$
\frac{\dot S}{k_B}= \lim_{(t_f-t_i)\to \infty}\, \frac{1}{t_f-t_i}\left\langle \log\left( \frac{\mathcal{P}_F}{\mathcal{P}_R} \right) \right\rangle=
 \lim_{(t_f-t_i)\to \infty}\,  \frac{1}{t_f-t_i}  \left\langle \frac{mg}{k_B T_0} \sum_{ij} \mathbb{K}_{ij}\int_{t_i}^{t_f}\!\!dt\,w_i(t)\,x_j(t) \right\rangle \,.
$$
Using the ergodicity, i.e. the equivalence of the time average and the average over the realization of the noise we conclude that
\begin{equation}
\dot S= \frac{mg}{  T_0} \sum_{ij} \mathbb{K}_{ij}  \left\langle w_i(t) x_j(t) \right\rangle \,.
\label{eq:epr}
\end{equation}
To evaluate $\left\langle w_i x_j \right\rangle $, we use the expansion in eigenmodes, Eqs.\eqref{eq:x(t)_solution}, 
and the correlation $\langle Q_k(t)\,P_q(t) \rangle$ that can be derived from the solution \eqref{eq:solQk}
$$
\langle Q_k(t)\,P_q(t) \rangle = e^{-\mu_k t} \int_{-\infty}^t \!\!\!ds\;e^{\mu_k s}\langle P_k(s)\,P_q(t) \rangle
$$
where we exploited the independence between the noises $P_k(t)$ and $\phi_k(t)$ in Eq.\eqref{eq:solQk}.
The explict expression of $\langle P_k(s)\,P_q(t) \rangle$ is given by the stationary statistical properties 
of $w_i(i)$, 
\begin{equation}
\langle P_k(t)\,P_q(s) \rangle = \frac{k_B T_a}{m} \delta_{k q}\;e^{-|t-s|/\tau_a}. 
\label{eq:PPcorr}
\end{equation}
After substituting Eq.\eqref{eq:PPcorr} into the above expression and performing the simple integration, we obtain the final formula 
\begin{equation}
\langle Q_k(t)\,P_q(t) \rangle = \frac{k_B T_a}{m}  \;\dfrac{\tau_a}{1+ \mu_k \tau_a} \delta_{k q}\;,
\label{eq:QPcorrelator}
\end{equation}
Using Eq.~\eqref{eq:QPcorrelator}, we have  
\begin{equation}
\langle w_i(t)x_j(t)\rangle =  \frac{k_B T_a \tau_a}{m}  
\sum_{k} u_i(k)\,u_j(k)\;\dfrac{1}{1+ \mu_k \tau_a} 
\label{eq:corrWX}
\end{equation}
leading to the following formula for the entropy production
\begin{equation}
\dot S = k_B \frac{T_a }{  T_0} g \sum_{k} \dfrac{\tau_a}{1+ \mu_k \tau_a}  
	\sum_{ij} u_i(k)\, \mathbb{K}_{ij} u_j(k) \,.
\end{equation}
The properties of the matrix $\mathbb{K}_{ij}$ and the normalization of the eigenvectors $\sum_i u^2_i(k) = 1$, yield
$$
\sum_{i} u_i(k) \sum_{j} \mathbb{K}_{ij} u_j(k)= 3 \lambda_k \sum_{i} u_i(k) u_i(k) = 3 \lambda_k 
$$
where the factor $3$ stems from the degeneracy of the eigenvalues.
Finally, we can write:
\begin{equation}
\dot S=3 k_B\frac{T_a }{T_0}  \sum_{k} \dfrac{g \lambda_k  \tau_a}{1+ \mu_k \tau_a}  =3k_B \gamma \frac{  T_a }{  T_0}  \sum_{k} \dfrac{\mu_k  \tau_a}{1+ \mu_k \tau_a} \,.
\label{eq:epr2}
\end{equation}
Since the eigenvalues are all positive, the EPR is positive as expected.

\section{Correlation functions derivation
\label{app:correla}}
This appendix shows the derivation of formulas~\eqref{eq:corrGk} and~\eqref{eq:Gk} needed to compute the residue-residue correlation matrix of the aGNM. The central quantity is the correlator $\langle Q_k(t)\,Q_p(t')\rangle$ among the modes, which, by using the solution~\eqref{eq:solQk}, can be written as 
$$
\langle Q_k(t)\,Q_p(t')\rangle  = e^{-(\mu_k t+\mu_q t')} \int_{-\infty}^t dt_1 \int_{-\infty}^{t'} dt_2\,e^{\mu_k t_1+\mu_p t_2}  \bigg[\langle P_k(t_1) P_p(t_2) \rangle + \frac{2 k_B T_0}{m\gamma}\langle \phi_k(t_1)\phi_p(t_2) \rangle  \bigg]\,,
$$
we defined for shortness $\mu_k = g\lambda_k/\gamma$. It is convenient to proceed by splitting the calculus in the active and thermal components. The active component of correlation is
$$
\langle Q_k(t)\,Q_p(t')\rangle_A  = e^{-(\mu_k t +\mu_q t')} \int_{-\infty}^t dt_1 \int_{-\infty}^{t'} dt_2\; e^{\mu_k t_1+\mu_p t_2} \langle P_k(t_1) P_p(t_2) \rangle\,, 
$$
where, in turn, the correlator of $P_k(t)$ can be derived directly from the stationary properties of the active noise, Eq.~\eqref{eq:evol2},
\begin{equation}
\langle P_k(t) P_p(t') \rangle = \delta_{kq}  \frac{k_B T_a}{m} \exp\{-|t-t'|/\tau_a\}\,.   
\end{equation}
Thanks to the presence of $\delta_{kq}$, this correlator becomes diagonal 
$$
\langle Q_k(t)\,Q_p(t')\rangle_A  =  \frac{k_B T_a}{m} \delta_{kq}\;e^{-\mu_k(t + t')} \int_{-\infty}^t dt_1 \int_{-\infty}^{t'} dt_2\;e^{\mu_k(t_1 + t_2)} e^{-|t_1-t_2|/\tau_a}\,.
$$
The double integral can be easily solved, leading to the expression
\begin{equation}
\langle Q_k(t)\,Q_p(t')\rangle_A  =  \frac{k_B T_a}{m}  \delta_{kq}\; \frac{\tau_a^2}{\mu_k\tau_a (\mu_k^2\tau_a^2-1)} \left(\mu_k\tau_a e^{-|t-t'|/\tau_a} -e^{-\mu_k |t-t'|}\right)\;.
\label{eq:Act}
\end{equation}
Likewise, the expression of the correlations between thermal modes reads
$$
\langle Q_k(t)\,Q_p(t')\rangle_T  = \frac{2 k_B T_0}{m\gamma} e^{-(\mu_k t+\mu_p t')} \int_{-\infty}^t dt_1 \int_{-\infty}^{t'} dt_2\;e^{\mu_k t_1+\mu_p t_2)} \langle \phi_k(t_1)\phi_p(t_2) \rangle 
$$
and taking into account that the delta-correlation of Brownian noise $\boldsymbol{\xi}$  also implies the delta-correlation of $\phi_k$, $\langle \phi_k(s)\phi_p(s')\rangle = \delta_{k,p} \delta(s-s')$, we have  
$$
\langle Q_k(t)\,Q_p(t')\rangle_T = \frac{2 k_B T_0}{m\gamma} \delta_{kp}\; e^{-\mu_k(t+t')} \int_{-\infty}^t dt_1 \int_{-\infty}^{t'} dt_2\;e^{\mu_k(t_1+t_2)}\delta(t_1-t_2) \,.
$$
The presence of the Dirac-delta simplifies the integral into 
$$
\langle Q_k(t)\,Q_p(t')\rangle_T = \frac{2 k_B T_0}{m\gamma} \delta_{kp}\ e^{-\mu_k(t+t')} \int_{-\infty}^{\mathrm{min}(t,t')} \!\!\!ds\; e^{2\mu_k s}
$$
leading to the following result for thermal modes correlation
\begin{equation}
\langle Q_k(t)\,Q_p(t')\rangle_T = \frac{k_B T_0}{m\gamma} \delta_{kp}\; \dfrac{e^{-\mu_k|t-t'|}}{\mu_k}\,.
\label{eq:Therm}
\end{equation}
Finally, combining expressions~\eqref{eq:Therm} and~\eqref{eq:Act}, we obtain the full expression for the mode-mode correlator
$$
\langle Q_k(t) Q_p(t')\rangle = \delta_{kq} \frac{k_B T_0}{m\gamma} \bigg[
\dfrac{e^{-\mu_k|t-t'|}}{\mu_k} + \frac{T_a}{T_0}\;
\frac{\gamma\tau_a^2}{\mu_k\tau_a (\mu_k^2\tau_a^2-1)} \left(\mu_k\tau_a e^{-|t-t'|/\tau_a} -e^{-\mu_k |t-t'|}\right)\bigg]
$$
that is needed to compute the residue-residue correlation matrix~\eqref{eq:corr}. The diagonal structure of the correlator in the mode indices confirms that mode independence is preserved, while their individual statistical or dynamical properties are altered.



\bibliographystyle{rsc} 

\bibliography{sample.bib} 

\end{document}